%% file: main.tex
\documentclass[acmsmall]{acmart}

\usepackage{amsmath}
\usepackage{array}
\usepackage[skip=1pt,labelfont=bf]{caption}
\usepackage{graphicx}
\usepackage{longtable}

\usepackage{hyperref}
\usepackage[ruled,vlined,lined,commentsnumbered]{algorithm2e}
\usepackage{amsfonts}
\usepackage{color, colortbl}
\usepackage{enumitem}
\usepackage{fancybox}
\usepackage{framed}
\usepackage{fontenc}
\usepackage{listings}
\usepackage{lscape}
\usepackage{makecell}
\usepackage{multirow}
\usepackage{pifont}
\usepackage{rotating}
\usepackage{setspace}
\usepackage{subfigure}
\usepackage{url}
\usepackage{soul}
\usepackage{textcomp}
\usepackage{xcolor}
\usepackage{wrapfig}
\usepackage{balance}
\usepackage{microtype}

\newcolumntype{L}[1]{>{\raggedright\let\newline\\\arraybackslash\hspace{0pt}}m{#1}}
\newcolumntype{C}[1]{>{\centering\let\newline\\\arraybackslash\hspace{0pt}}m{#1}}
\newcolumntype{R}[1]{>{\raggedleft\let\newline\\\arraybackslash\hspace{0pt}}m{#1}}

\newboolean{showcomments}
\setboolean{showcomments}{true}
\ifthenelse{\boolean{showcomments}}
 { \newcommand{\mynote}[2]{
      \fbox{\bfseries\sffamily\scriptsize#1}
        {\small$\blacktriangleright$\textsf{\emph{#2}}$\blacktriangleleft$}}}
        { \newcommand{\mynote}[2]{}}

\newcommand{\toolname}{CodeGPTSensor\xspace}

\newcommand{\datasetname}{{HMCorp}\xspace}

\newcommand{\code}[1]{\texttt{\small #1}}

\definecolor{lightgray}{gray}{0.9}
\definecolor{graybg}{gray}{0.9}
\definecolor{grayframe}{gray}{0.5}

\newcommand{\intuition}[1]{
\cornersize{0.2}
\setlength{\fboxsep}{5pt}
\noindent
\Ovalbox{
\parbox{0.96\linewidth}{
    \em {#1}
}
}
}

\newcommand{\promptbox}[1]{
\colorbox{graybg}{
    \parbox{0.9\linewidth}{
        \noindent \em \textbf{Prompt:} \\ #1
    }
}
}

\makeatletter
\DeclareRobustCommand\onedot{\futurelet\@let@token\@onedot}
\def\@onedot{\ifx\@let@token.\else.\null\fi\xspace}

\def\eg{{e.g}\onedot} 
\def\ie{{i.e}\onedot}

\def\etal{{et al}\onedot}
\makeatother

\AtBeginDocument{%
  }

\setcopyright{acmlicensed}
\copyrightyear{2024}
\acmYear{2024}
\acmDOI{XXXXXXX.XXXXXXX}

\acmJournal{TOSEM}
\begin{document}

\title{Distinguishing LLM-generated from Human-written Code by Contrastive Learning}

\author{Xiaodan Xu}
\affiliation{%
  \institution{State Key Laboratory of Blockchain and Data Security, Zhejiang University}
  \city{Hangzhou}
  \country{China}
  % \postcode{78229}
  }
  % \affiliation{%
  % \institution{Hangzhou High-Tech Zone (Binjiang) Blockchain and Data Security Research Institute}
  % \city{Hangzhou}
  % \country{China}
  % }
\email{xiaodanxu@zju.edu.cn}

\author{Chao Ni}
\authornote{This is the corresponding author.\\Chao Ni is also with Hangzhou High-Tech Zone (Binjiang) Blockchain and Data Security Research Institute, Hangzhou, China.}
\affiliation{%
  \institution{State Key Laboratory of Blockchain and Data Security, Zhejiang University}
  % \streetaddress{8600 Datapoint Drive}
  \city{Hangzhou}
  % \state{Texas}
  \country{China}
  % \postcode{78229}
  }
  % \affiliation{%
  % \institution{Hangzhou High-Tech Zone (Binjiang) Blockchain and Data Security Research Institute}
  % % \streetaddress{8600 Datapoint Drive}
  % \city{Hangzhou}
  % % \state{Texas}
  % \country{China}
  % % \postcode{78229}
  % }
\email{chaoni@zju.edu.cn}

\author{Xinrong Guo}
\affiliation{%
  \institution{State Key Laboratory of Blockchain and Data Security, Zhejiang University}
  % \streetaddress{8600 Datapoint Drive}
  \city{Hangzhou}
  % \state{Texas}
  \country{China}
  % \postcode{78229}
  }
\email{22151096@zju.edu.cn}

\author{Shaoxuan Liu}
\affiliation{%
  \institution{State Key Laboratory of Blockchain and Data Security, Zhejiang University}
  % \streetaddress{8600 Datapoint Drive}
  \city{Hangzhou}
  % \state{Texas}
  \country{China}
  % \postcode{78229}
  }
\email{22221076@zju.edu.cn}

 \author{Xiaoya Wang}
\affiliation{%
  \institution{State Key Laboratory of Blockchain and Data Security, Zhejiang University}
  % \streetaddress{8600 Datapoint Drive}
  \city{Hangzhou}
  % \state{Texas}
  \country{China}
  % \postcode{78229}
  }
\email{22151096@zju.edu.cn}

\author{Kui Liu}
% \authornotemark[3]
\affiliation{%
  \institution{Software Engineering Application Technology Lab, Huawei}
  \city{Hangzhou}
  \country{China}
  }
\email{brucekuiliu@gmail.com}

\author{Xiaohu Yang}
\affiliation{%
  \institution{State Key Laboratory of Blockchain and Data Security, Zhejiang University}
  % \streetaddress{8600 Datapoint Drive}
  \city{Hangzhou}
  % \state{Texas}
  \country{China}
  % \postcode{78229}
  }
\email{yangxh@zju.edu.cn}

% \renewcommand{\shortauthors}{Xiaodan Xu et al.}

\begin{abstract}
Large language models (LLMs), such as ChatGPT released by OpenAI, have attracted significant attention from both industry and academia due to their demonstrated ability to generate high-quality content for various tasks.
Despite the impressive capabilities of LLMs, there are growing concerns regarding their potential risks in various fields, such as news, education, and software engineering.
Recently, several commercial and open-source LLM-generated content detectors have been proposed, which, however, are primarily designed for detecting natural language content without considering the specific characteristics of program code.
This paper aims to fill this gap by proposing a novel ChatGPT-generated code detector, \toolname, based on a contrastive learning framework and a semantic encoder built with UniXcoder. 
To assess the effectiveness of \toolname on differentiating ChatGPT-generated code from human-written code, we first curate a large-scale Human and Machine comparison Corpus (\datasetname), which includes 550K pairs of human-written and ChatGPT-generated code (\ie, 288K Python code pairs and 222K Java code pairs). 
Based on the \datasetname dataset, our qualitative and quantitative analysis of the characteristics of ChatGPT-generated code reveals the challenge and opportunity of distinguishing ChatGPT-generated code from human-written code with their representative features.
Our experimental results indicate that \toolname can effectively identify ChatGPT-generated code, outperforming all selected baselines.
\end{abstract}




\received{18 October 2023}
\received[revised]{14 June 2024}
\received[revised]{5 September 2024}
\received[accepted]{14 October 2024}

\maketitle

\input{sections/introduction}
\label{sec:introduction}

\input{sections/motivation}

\label{sec:motivation}

\input{sections/approach}

\input{sections/experiment_design}

\input{sections/experiment_results}

\input{sections/discussion}

\input{sections/threats_to_validate}
\label{sec:threats}

\input{sections/related_work}

\input{sections/conclusion}
\label{sec:conclusion}

\begin{acks}
This work was supported by the National Natural Science Foundation of China (Grant No.62202419), the Fundamental Research Funds for the Central Universities (No. 226-2022-00064),
Zhejiang Provincial Natural Science Foundation of China (No. LY24F020008),
the Ningbo Natural Science Foundation (No. 2022J184), 
and the State Street Zhejiang University Technology Center.
\end{acks}

\bibliographystyle{ACM-Reference-Format}
\bibliography{main}

% \appendix

\end{document}

%% file: sections/introduction.tex
\section{Introduction}

As large language models (LLMs) continue to advance, the quality of LLM-generated content has significantly improved.
As the milestone of LLMs, ChatGPT~\cite{openaichatgpt}, developed by OpenAI, has attracted wide attention and sparked extensive discussions in academia and industry.
ChatGPT is fine-tuned from a model in the GPT-3.5 series using Reinforcement Learning from Human Feedback (RLHF)\footnote{\url{https://openai.com/blog/chatgpt}}
, which demonstrates the capacity to understand user questions, comprehend conversational context, and generate coherent responses, as evidenced by its evaluations across a wide range of domains, including story-writing~\cite{yao2019plan}, medical education~\cite{kung2023performance} and code generation~\cite{chen2021evaluating}.

There has been a growing number of research on LLMs for software engineering tasks. In particular, LLMs for automated code generation, exemplified by Codex~\cite{chen2021evaluating}, have emerged as promising tools to improve coding productivity and efficiency~\cite{xu2022systematic}.
General text generation LLMs like ChatGPT~\cite{openaichatgpt} have also demonstrated promising performance in generating code. 
Despite the considerable benefits, surprisingly, a sentiment analysis study shows that fear is the dominant emotion people feel about the code generation capabilities of ChatGPT~\cite{feng2023investigating}. Concerns have been raised about how LLMs for automated code generation impact software engineering, the programming community, and education.
For example, Stack Overflow, a famous online community for developers, has banned contributions by ChatGPT as a precaution against ChatGPT-generated unreliable content that may shatter the community's long-established trust~\cite{stackoverflowbangpt}.
Several studies in the education domain report that students have already begun to use ChatGPT to complete their class assignments or even cheat on exams~\cite{nietzel2023more, susnjak2022chatgpt}.
Applying LLMs for code generation also raises concerns in the software industry, particularly regarding ethical and code quality problems.
For example, Copilot, which is built on Codex, is questioned with code plagiarism\footnote{\url{https://twitter.com/mitsuhiko/status/1410886329924194309}} and generating vulnerable code~\cite{copilotflaw}.
Liu et al.~\cite{liu2023refining} find that ChatGPT-generated code often faces quality issues like code style and maintainability problems, wrong outputs, compilation and runtime errors, and performance inefficiencies. 
Zhong et al.~\cite{zhong2023study} find that LLM-generated code often contains API misuses.
Several studies demonstrate that ChatGPT often produces insecure code that exhibits vulnerabilities~\cite{khoury2023secure, liu2023no}.

One possible solution to mitigate the above issues is to develop a technique that predicts how likely a given piece of code is generated by LLM so that LLM-generated code can be differentiated from human-written code.
This technique can help community administrators block answers with code snippets that are likely to be generated by LLMs.
Besides, the estimated probability can also be a reference factor for evaluating students' scores on programming assignments and exams. 
Since the use of LLM-generated code in real software projects involves concerns related to provenance, security, and maintainability, it is also important to distinguish LLM-generated code during code review and quality assurance processes. 
For example, code commits that are likely to contain code generated by LLMs should be prioritized for inspection and given extra attention.
However, existing AI-generated content (AIGC) detectors primarily focus on detecting natural language content generated by LLMs, such as prose and news articles~\cite{gpt2detector,mitchell2023detectgpt, robertaqa,chatgptzero,copyleaks,writer,aitextclassifier}.
These AIGC detectors have not considered the fundamental differences between code and natural language text and have been reported to be unreliable in detecting AI-generated code in a recent empirical study~\cite{pan2024assessing}.
To fill this gap, we aim to explore the technology to address the challenge of distinguishing between LLM-generated and human-written code.

Inspired by the recent success of contrastive learning in code-related tasks such as code clone detection~\cite{jain2020contrastive}, code retrieval and summarization~\cite{bui2021self}, in this paper, we propose a novel ChatGPT-generated code detection method, \textbf{\toolname}, based on the contrastive learning framework~\cite{hoffer2015deep} with the code semantic encoder UniXcoder~\cite{guo2022unixcoder}. 
UniXcoder is a pre-trained model that incorporates semantic and syntax information from the source code to support code-related tasks.
To learn distinct code representations for differentiating ChatGPT-generated code from human-written code, we utilize a contrastive learning framework to fine-tune UniXcoder, which ensures that pairs of samples with the same label exhibit smaller distances in the embedding space compared to pairs with different labels.

To investigate the effectiveness of \toolname on differentiating ChatGPT-generated code from human-written code, 
we first curate a dataset with \textbf{550K} pairs of human-written and ChatGPT-generated code (\ie, 288K Python code pairs and 222K Java code pairs), namely the \textbf{Human and Machine comparison Corpus (\datasetname)}.
The human-written samples in \datasetname are sourced from functions with corresponding docstrings collected from about 17K real-world projects on GitHub. 
Each human-written function is paired with a function generated by ChatGPT based on the corresponding docstring.
With our curated dataset, we first conduct a human study with developers to investigate whether they can manually distinguish between AI-generated code and human-written code. The results reveal that the differentiating task is a big challenge for developers without any assistant tool, where their decisions mainly rely on coding style, code logic, and intuition based on their knowledge.
Following that, we make a qualitative and quantitative analysis of the characteristics of ChatGPT-generated code against human-written code.
Eventually, we conduct large-scale experiments on \datasetname to assess the performance of \toolname in detecting ChatGPT-generated code and compare it against existing AIGC detectors, including two commercial tools, two zero-shot methods, and four training-based methods.
Experimental results indicate that \toolname outperforms the selected approaches in effectively detecting code generated by ChatGPT. 
In brief, the main contributions of this paper are summarized as follows:
\begin{itemize}[leftmargin=*]
    \item We curate a large-scale dataset named \datasetname, which contains 288K Python and 222K Java pairs of human-written and ChatGPT-generated code snippets.
    \item We conduct a human study on manually identifying ChatGPT-generated code, followed by quantitative and qualitative analysis to discern the differences between ChatGPT-generated and human-written code, which reveals several characteristics inherent to ChatGPT-generated code.
    \item We propose \toolname, a contrastive learning-based approach of the first to detect code generated by LLMs.
    To facilitate future research in LLM-generated code detection, we have open-sourced our dataset and source code~\cite{gitreplication}.
\end{itemize}

% The rest of this paper is organized as follows. 
% Section~\ref{sec:motivation} first introduces the motivation of our work.
% Following that, Section~\ref{sec:approach} introduces the design of \toolname. 
% Section~\ref{sec:design} presents the experimental setting including the constructed dataset, compared baselines, the implementation details, the research question, and considered performance measures.
% Section~\ref{sec:results} reports the experimental results.
% Following that, some threats to validity are presented in Section~\ref{sec:threats}. 
% Section~\ref{sec:related_work} describes related prior work.
% Finally,  we conclude our work and mention future plans in Section~\ref{sec:conclusion}.

%% file: sections/motivation.tex
\section{Motivation}

\subsection{Challenges \& Opportunities}
\label{sec:motivating_example}
AI-generated code can exhibit high semantic and syntactic similarity to human-written code, making distinguishing between them a significant challenge.
However, subtle differences in the use of code syntax and structure, etc., may provide opportunities to address this challenge.

\begin{figure}[!htbp]
\centerline{
    \includegraphics[width=.68\linewidth]{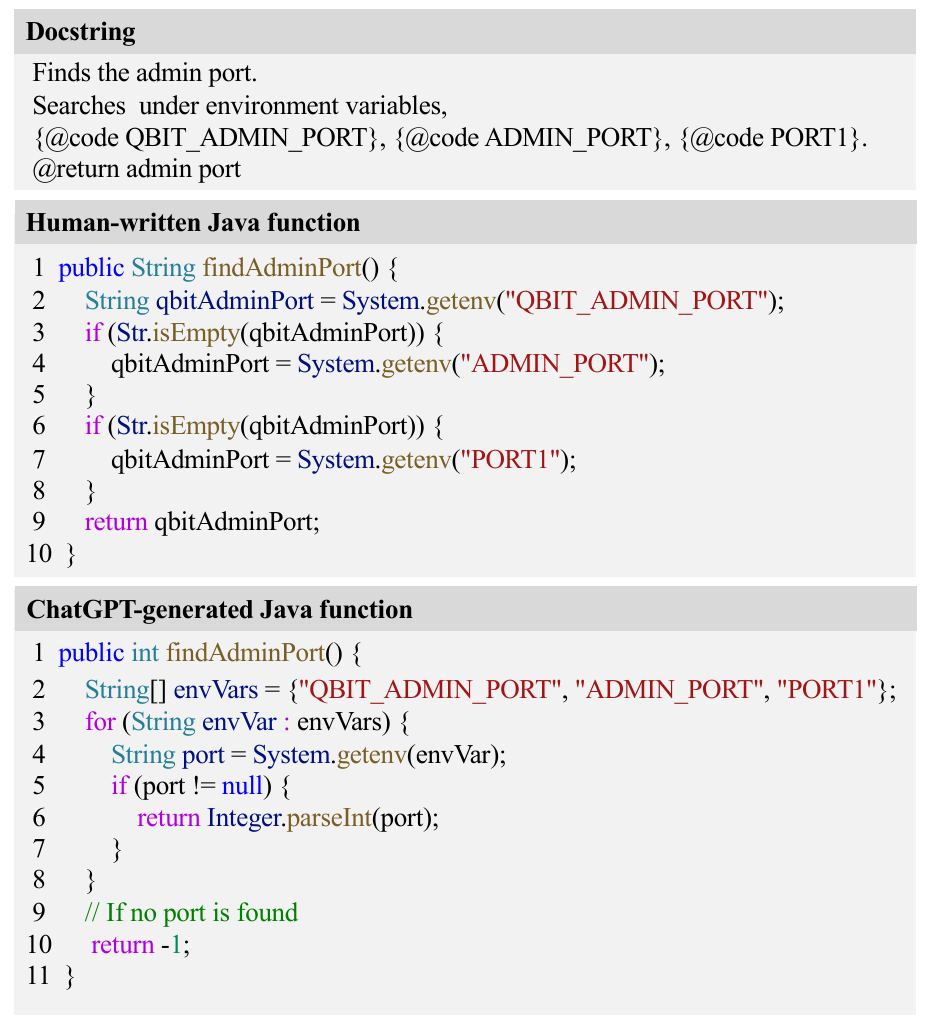}
    }
    \caption{An example pair of human-written and ChatGPT-generated Java functions from \datasetname-gj270412~\cite{javaexample}.
    }
    \Description{A human-written Java function with its docstring, paired by a ChatGPT-generated Java function that is very similar to the human-written one.}
    \label{fig:java_example}
\end{figure}

\textbf{\textit{Semantically similar Code}:}
Figure~\ref{fig:java_example} shows an example pair of Java functions in the \datasetname dataset (details of the \datasetname dataset will be presented in Section~\ref{sec:dataset}), where the first one is written by a developer, and the other is generated by ChatGPT based on the corresponding function description text (\ie, Docstring).
% In line 1, both functions start with the same method signature (i.e., access modifier, method name, and parameter) except for the return data type.
% Subsequently, the two functions are implemented in different ways; however, both share the core implementation of invoking \code{System.getenv()}\footnote{A Java method that takes in a string variable and returns the corresponding string value for the environment variable if it exists, or returns null otherwise.} to get the related port.
Semantically, both functions basically do the same thing: search the port with three names (\code{QBIT\_ADMIN\_PORT}, \code{ADMIN\_PORT} and \code{PORT1}) one by one by invoking \code{System.getenv()} until an environment variable is retrieved.
Syntactically, however, the functions are implemented in two different ways.
First of all, the human-written function directly returns the \code{String} value returned by \code{System.getenv()}, while the ChatGPT-generated function converts the string value returned by \code{System.getenv()} into the integer type \code{int}, and uses \code{return -1;} to explicitly return failure (i.e., no port is found).
Besides, the human-written function creates branches with two subsequent \code{if} statements,
while ChatGPT-generated function uses a \code{for} loop with a nested \code{if} statement.
In this example, the human-written function is easier to understand for the readers with its explicit handling of three distinct cases for the three ports, while the ChatGPT-generated function considers a more concise invocation of the API \code{System.getenv()} with a bit more complicated structure.

\addvspace{0.2em}
\intuition{\textbf{Observation 1:} AI-generated functions can perform functionalities semantically similar to human-written ones, which makes differentiation challenging; however, their different implementing ways and coding styles may provide the opportunity to differentiate them.}

\begin{figure}[!htbp]
\centerline{
    \includegraphics[width=.7\linewidth]{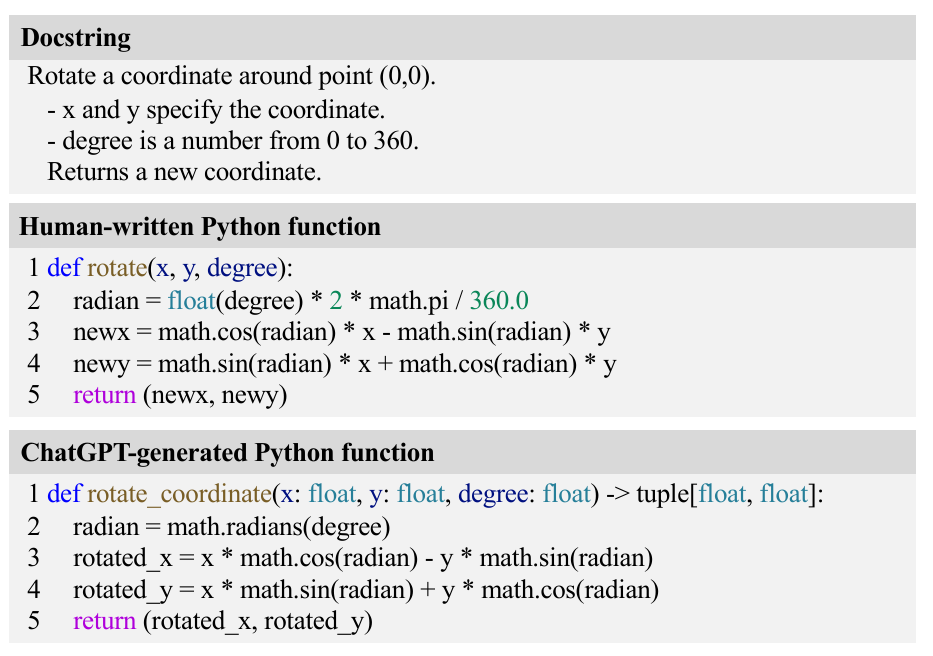}
    }
    \caption{An example pair of human-written and ChatGPT-generated Python functions from \datasetname-gp120870~\cite{pythonexample}.}
    \label{fig:python_example}
    \Description{A human-written Python function with its docstring, paired by a ChatGPT-generated Python function that is very similar to the human-written one.}
\end{figure}

\textbf{\textit{Syntactically similar Code}:}
Figure~\ref{fig:python_example} shows an example of a human-written and ChatGPT-generated Python code pair that exhibits semantic and syntactic similarity.
The two functions are extremely similar from line 3 to line 5 for their core implementation of rotating the coordinate, except for minor layout differences (i.e., the multiplier and multiplicand changes place, \eg, \code{math.cos(radian) * x} and \code{x * math.cos(radian)}) and lexical differences (i.e., different local variable names, \eg, \code{newx} and \code{rotated\_x}).
The two functions also differ in their function names, with one using a simple verb \code{rotate} and the other using a verb-object construction \code{rotate\_coordinate}, and differ in how they calculate \code{radian}, where the former explicitly writes out the formula while the latter employs a more concise approach by invoking \code{math.radians}.
Additionally, the ChatGPT-generated function explicitly specifies the type of the parameters and return values.
Both examples in this section imply that ChatGPT-generated code more strictly adheres to certain coding standards and rules (e.g., favoring more concise implementations and meaningful local variable names).
Such context differences could be the key to distinguishing semantically and syntactically similar AI-generated and human-written code.

\addvspace{0.2em}
\intuition{
\textbf{Observation 2:} 
\textit{Human-written and AI-generated code can exhibit significant semantic and syntactic similarity, with only subtle differences in their context, which poses a new challenge for the automatic detection of AI-generated code.}
}

\subsection{Preliminary Study}
At the beginning of this work, we conduct a preliminary study to check whether the six existing AIGC detectors (cf. Section~\ref{baselines}), including two commercial detectors Writer~\cite{writer} and ZeroGPT~\cite{zerogpt}, the state-of-the-art zero-shot detector DetectGPT~\cite{mitchell2023detectgpt}, and three RoBERTa-based detectors RoBERTa-single~\cite{robertasingle}, RoBERTa-QA~\cite{robertaqa} and OpenAI GPT-2 Output Detector (\ie, GPT2-detector)~\cite{gpt2detector}, % state-of-the-art AIGC detectors (\ie, commercial detectors Writer~\cite{writer} and ZeroGPT~\cite{zerogpt}, the SOTA zero-shot detector DetectGPT~\cite{mitchell2023detectgpt}, and RoBERTa-based detectors RoBERTa-single~\cite{robertasingle}, RoBERTa-QA~\cite{robertaqa} and GPT2-detector~\cite{gpt2detector}, cf. Section~\ref{baselines}) 
can distinguish between human-written and ChatGPT-generated code. Taking the two aforementioned code pairs as examples for illustration, we summarized the detection results in Table~\ref{tab:motivation_result}.
% identify the differences between human-written code and ChatGPT-written code with the aforementioned two examples.
% \xxd{
% Table~\ref{tab:motivation_result} shows the detection results.
Overall, these detectors exhibit a notable tendency to misidentify AI-generated code as ``Human-written''.
Meanwhile, they misclassify human-written code as ``AI-generated'' in about half of the cases.
In particular, DetectGPT labels all code snippets as ``AI-generated'', while RoBERTa-single and RoBERTa-QA label all code snippets as ``Human-written''.
These findings highlight the challenges current AIGC detectors face in effectively distinguishing between human-written and ChatGPT-generated code and underscore the urgent need for more sophisticated methods to enhance the accuracy of ChatGPT-generated code detection.
To address this challenge, in this work, we aim to seek an effective methodology by considering the specific characteristics of program code and learning useful code representations to separate human-written and AI-generated code.

\begin{table}[t]
  \centering
  \caption{Detection results for the Java and Python examples from AI content detectors. Correct answers are highlighted in green, while wrong answers are highlighted in red.}
  \resizebox{0.65\linewidth}{!}
  {
    \begin{tabular}{lcccc}
    \toprule
    \multirow{2}[2]{*}{\textbf{Detector}} & \multicolumn{2}{c}{\textbf{Java (ID-gj270412)}} & \multicolumn{2}{c}{\textbf{Python (ID-gp120870)}} \\
          & \textbf{ChatGPT (AI)} & \textbf{Human} & \textbf{ChatGPT (AI)} & \textbf{Human} \\
    \midrule
Writer~\cite{writer} & \cellcolor[rgb]{ 1,  .863,  .847}Human & \cellcolor[rgb]{ 1,  .863,  .847}AI & \cellcolor[rgb]{ 1,  .863,  .847}Human & \cellcolor[rgb]{ .882,  .965,  .729}Human \\
ZeroGPT~\cite{zerogpt} & \cellcolor[rgb]{ 1,  .863,  .847}Human & \cellcolor[rgb]{ .882,  .965,  .729}Human & \cellcolor[rgb]{ .882,  .965,  .729}AI & \cellcolor[rgb]{ 1,  .863,  .847}AI \\
DetectGPT~\cite{mitchell2023detectgpt} & \cellcolor[rgb]{ .882,  .965,  .729}AI & \cellcolor[rgb]{ 1,  .863,  .847}AI & \cellcolor[rgb]{ .882,  .965,  .729}AI & \cellcolor[rgb]{ 1,  .863,  .847}AI \\
RoBERTa-single~\cite{robertasingle} & \cellcolor[rgb]{ 1,  .863,  .847}Human & \cellcolor[rgb]{ .882,  .965,  .729}Human & \cellcolor[rgb]{ 1,  .863,  .847}Human & \cellcolor[rgb]{ .882,  .965,  .729}Human \\
RoBERTa-QA~\cite{robertaqa} & \cellcolor[rgb]{ 1,  .863,  .847}Human & \cellcolor[rgb]{ .882,  .965,  .729}Human & \cellcolor[rgb]{ 1,  .863,  .847}Human & \cellcolor[rgb]{ .882,  .965,  .729}Human \\
GPT2-detector~\cite{gpt2detector} & \cellcolor[rgb]{ 1,  .863,  .847}Human & \cellcolor[rgb]{ 1,  .863,  .847}AI & \cellcolor[rgb]{ 1,  .863,  .847}Human & \cellcolor[rgb]{ .882,  .965,  .729}Human \\
    \bottomrule
    \end{tabular}%
  }
  \label{tab:motivation_result}%
\end{table}%

%% file: sections/approach.tex
\section{Approach}
\label{sec:approach}

\begin{figure*}[t]
\centerline{
    \includegraphics[width=\linewidth]{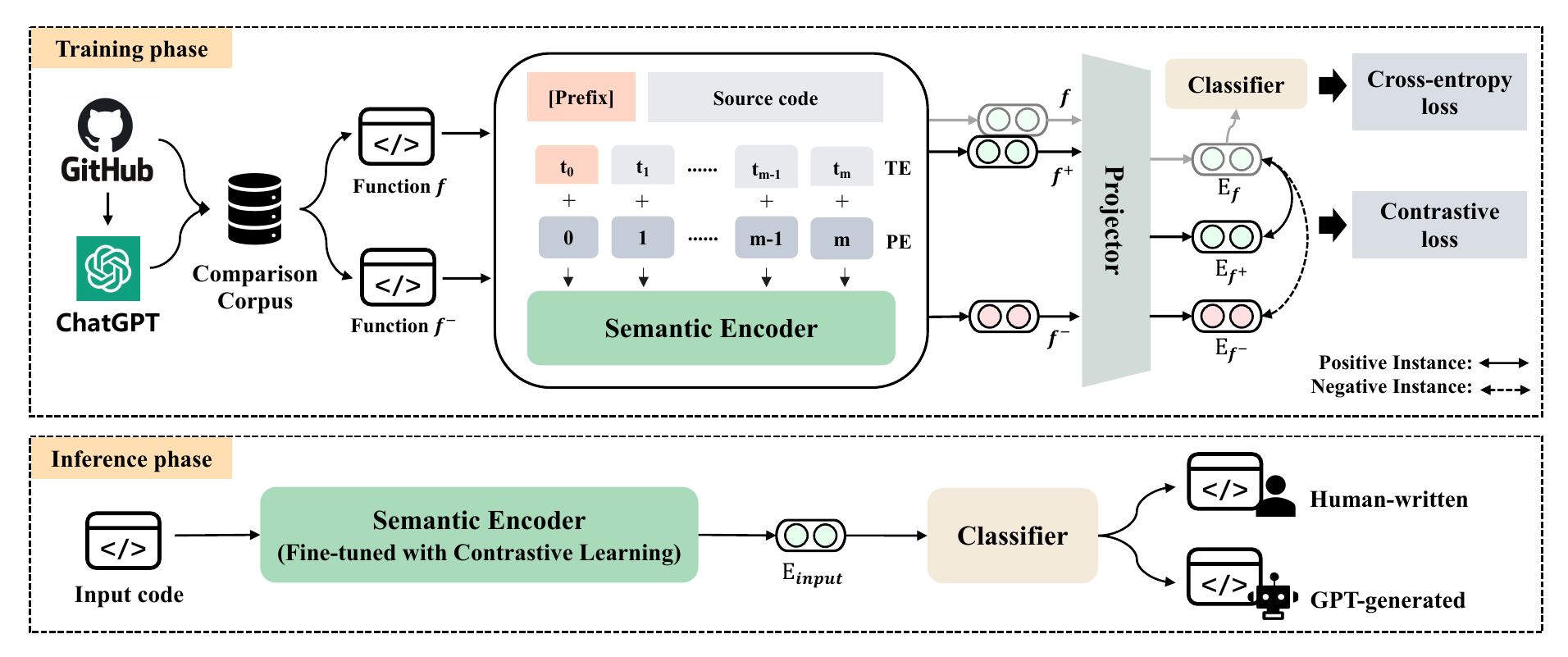}
    }
    \caption{The framework of \toolname.}
    \Description{The framework of \toolname, which consists of the training phase and the inference phase. In the training phase, UniXcoder is fine-tuned using contrastive learning for detecting LLM-generated code.}
    \label{fig:framework}
\end{figure*}

This section introduces our model \toolname for ChatGPT-generated code detection. As illustrated in Figure~\ref{fig:framework}, \toolname consists of two phases: online training and offline inference.
During the training phase, \toolname first utilizes UniXcoder as the semantic encoder to embed the function source code into preliminary semantic representations (i.e., hidden vectors).
To learn class-separation features, the contrastive learning framework is leveraged to fine-tune UniXcoder by minimizing the distance between functions of the same class while separating functions from different classes. 
After training, a given code snippet can be classified by \toolname as either human-written or ChatGPT-generated during the inference phase.

\subsection{Semantic Encoder with UniXcoder}

Program code can be encoded into token sequences similarly to how natural language text is processed in deep learning.
However, program code has distinct characteristics, including highly structured syntax, unique naming conventions, task-specific and non-word identifiers, etc.
In the literature, various code intelligence tasks (\eg, defect prediction~\cite{ni2022best}, vulnerability detection~\cite{fu2022linevul} and code summarization~\cite{zhu2019automatic}) have achieved promising results by encoding the specific characteristics of program code using UniXcoder~\cite{guo2022unixcoder}.
UniXcoder is a unified pre-trained model for programming language based on Transformer~\cite{vaswani2017attention} that can incorporate semantic and syntax information from both code comment and Abstract Syntax Tree (AST).
The model uses mask attention matrices with prefix adapters to control its behavior (\ie, $[\mathit{Enc}]$ for encoder-only mode, $[\mathit{Dec}]$ for decoder-only mode, and $[\mathit{E2D}]$ for encoder-decoder mode). 
Given an input function, UniXcoder first transforms its AST into a sequence that retains all structural information from the tree. Then, the flattened AST sequence is concatenated with a prefix as an input token sequence. 

Inspired by the successful applications of UniXcoder in tasks from the software engineering domain, we adopt UniXcoder as the basic semantic encoder in \toolname to obtain the code representations of the input functions. 
For \toolname, a binary classification model to predict whether the given code is ChatGPT-generated, we set the prefix as $[\mathit{Enc}]$ to use the encoder-only mode of UniXcoder and fine-tune it on our dataset. 
Following Guo et al.~\cite{guo2022unixcoder}, we only keep leaves of AST (i.e., source code) as input during the fine-tuning phase.

\subsection{Contrastive Learning}

As presented in Section~\ref{sec:motivating_example}, human-written and AI-generated code can exhibit significant semantic and syntactic similarities, making it challenging to distinguish them. The key obstacle lies in capturing their representative features from the subtle differences in their implementation ways and contexts.
In this work, we thus propose to minimize distances between code representations within the same class and maximize distances between code representations of different classes (i.e., human-written or ChatGPT-generated) with contrastive learning.

Hoffer and Ailon~\cite{hoffer2015deep} propose the triplet network for contrastive learning.
In the context of function-level code-related tasks, the input of the network is a triplet $(f, f^+, f^-)$, where $f$ is a given function (\ie, the anchor), $f^+$ is its positive equivalent, and $f^-$ is the negative one which is of a different class from $f$. 
The triplet network can learn useful representations for the downstream classifying tasks by minimizing the distance between $f$ and $f^+$ while maximizing the distance between $f$ and $f^-$.
In the literature, contrastive learning has demonstrated its effectiveness in various code-related tasks, such as clone detection~\cite{jain2020contrastive} and code retrieval and summarization~\cite{bui2021self}.
Therefore, we adopt the triplet network-based contrastive learning approach to build our model. 
In this work, for a given function $f$, we assume that $f^+$ belongs to the same class as $f$, while $f^-$ belongs to a different class. 
Specifically, if $f$ represents a human-written function, $f^+$ will also be human-written, while $f^-$ will be ChatGPT-generated. 
\toolname uses UniXcoder as the base model to obtain corresponding embeddings of $f$, $f^+$, and $f^-$. 
For contrastive learning, we follow previous works~\cite{gao2021simcse, guo2022unixcoder} to forward the same input using different hidden dropout masks (dropout probability = 0.1) to generate $f$ and $f^+$.
Concretely, the original function $f$ is input twice into the semantic encoder (i.e., UniXcoder) applied with dropout to obtain two varying code representations for the same function, which can be interchangeably treated as positive instances for contrastive learning~\cite{hoffer2015deep}. After that, we transform the outputs from UniXcoder into final embeddings $E_f$, $E_{f^+}$, and $E_{f^-}$ using a structure called \textit{projector}.
Then, a contrastive loss computed based on the cosine distance of the embeddings is used to train the network.
Specifically, the training objective is to minimize the distance between $f$ and $f^+$ while separating $f$ and $f^-$, formulated as below:
\begin{equation*}
    max(||E_{f} - E_{f^+}|| - ||E_f - E_{f^-}|| + \epsilon, 0)
\end{equation*}
where $\epsilon$ represents the margin of the distance between $f$ and $f^-$, and is set to 1 by default.

Through contrastive learning, \toolname can learn useful code representation $E_f$ for an input function $f$. 
By further passing it to a classification layer and applying the softmax function, we can obtain the probability for each class label. A probability threshold of 0.5 is used as a cut-off value to indicate whether the function is human-written or ChatGPT-generated.

%% file: sections/experiment_design.tex
\section{Experiment Setup}
\label{sec:design}

This section introduces the construction process of the dataset named \datasetname, the selected baselines and their implementations, the research questions, and the evaluation metrics.

% \subsection{Curate Dataset} 
\subsection{Dataset Construction}
\label{sec:dataset}
ChatGPT has demonstrated its capability to automatically generate code when provided with requirements expressed in natural language~\cite{openaichatgpt}.
In this work, we want to investigate the distinct characteristics exhibited by ChatGPT-generated codes and assess how they differ from human-written ones.
Furthermore, we aim to explore the feasibility of distinguishing between human-written and ChatGPT-generated code with code semantic encoding and contrastive learning. To this end, we curate the first large-scale dataset comprising human-written and ChatGPT-generated code pairs, namely the \datasetname dataset, which is built with two steps: \ding{172} collecting code snippets and corresponding functional descriptions from real-world open-source software projects and \ding{173} employing ChatGPT to generate code based on the functional descriptions collected in the previous step.

\begin{table}[tbp]
  \centering
  \caption{Distribution of noise in the CodeSearchNet dataset.}
  \resizebox{0.4\linewidth}{!}
  {
    \begin{tabular}{rrr}
    \toprule
    \textbf{Type of Noisy Sample} & \textbf{Python} & \textbf{Java} \\
    \midrule
    HTML tags & 11,643  & 97,160  \\
    URL   & 14,276  & 11,411  \\
    Non-Literal & 6,336  & 22,739  \\
    Interrogation & 992   & 1,207  \\
    Under-Development & 5,562  & 17,964  \\
    Auto Code & 2,006  & 16,560  \\
    Empty Function & 340   & 126  \\
    \midrule
    \rowcolor{lightgray}\textit{Removed noisy samples} & 36,088  & 152,064  \\
    \bottomrule
    \end{tabular}%
  }
  \label{tab:remove_noise}%
\end{table}%

\begin{table}[tbp]
  \centering
  \caption{Preprocessing steps for the CodeSearchNet dataset.}
  \resizebox{0.75\linewidth}{!}
  {
    \begin{tabular}{lrrp{20em}}
    \toprule \multicolumn{1}{l}{\textbf{Dataset Type}}
          & \multicolumn{1}{l}{\textbf{Python}} & \multicolumn{1}{l}{\textbf{Java}} & \multicolumn{1}{l}{\textbf{Description}} \\
    \midrule
    {Original} & 457,461 & 496,688 & \multicolumn{1}{l}{The original CSN dataset} \\
    {Noise Free} & 421,373 & 344,624 & \multicolumn{1}{l}{After removing noisy samples} \\
    {Final} & 336,668 & 275,467 & ${len(docstring)} \in (10\% \mathit{quantile}, 90\% \mathit{quantile}]$
    \newline{} \ie, $(33, 614]$ for Python and $(38, 372]$ for Java \\
    \bottomrule
    \end{tabular}%
  }
  \label{tab:csn}%
\end{table}%

\subsubsection{Collecting Human-written Code} \label{sec:human-code-collection}
Collecting and pre-processing large amounts of human-written code with corresponding documentation from scratch is tedious and time-consuming. Therefore, we directly gather code written by software developers from the publicly available CodeSearchNet (CSN)~\cite{husain2019codesearchnet} dataset, which is sourced from open-source GitHub repositories and contains approximately 2 million pairs of documentation comments (\ie, docstrings) and the corresponding function code spanning six programming languages (i.e., Go, Java, JavaScript, PHP, Python, and Ruby).
We use code samples from CSN to represent human-written code snippets, which come from real-world software projects and are diverse in scale and functionalities.
Considering the popularity of Python and Java languages\cite{tiobeindex} as well as the usage limitation of ChatGPT, we collect $<$\textit{docstring, code}$>$ pairs in Python (457,461 pairs) and Java (496,688 pairs) languages from CSN to build our dataset. 
To improve the quality of the collected data, we filter out noisy samples from the Python and Java code pairs following the rules proposed by Shi \etal~\cite{shi2022we}. Specifically, we remove the following types of noisy samples:
\begin{itemize}
    \item \textit{HTML tags:} the docstring contains HTML tags. A docstring with HTML tags is prone to be auto-generated instead of human-written and could include noisy information.
    \item \textit{URL:} the docstring contains URLs. Developers may use URLs to refer to external references, which can potentially influence the subsequent code generation step, as it remains uncertain whether ChatGPT can leverage such external information when generating code based on the docstring.
    \item \textit{Non-Literal:} the docstring contains non-ASCII characters.
    \item \textit{Interrogation:} the docstring contains ``?'', ``what'', ``how'', etc. Interrogations are primarily used for communication instead of describing the functionalities.
    \item \textit{Under-Development Comments:} the docstring contains ``todo'', ``deprecate'', ``FIXME'', etc. Under-development comments are temporary notes for future development, which are inappropriate for text-to-code generation.
    \item \textit{Empty Function:} the method body is empty, or only contains a \textit{pass} statement (for Python). An unimplemented empty function is often not matched to its docstring semantically. 
    \item \textit{Auto code:} the code is auto-generated by IDEs (e.g., setter, getter, tester, etc.). The docstring of the auto-generated method is often similar to the method name and thus contains limited information on the functionality.
\end{itemize}

The distribution of noisy samples in CSN is shown in Table~\ref{tab:remove_noise}. Note that a single $<$\textit{docstring, code}$>$ pair may include multiple types of noise, and thus the sum of individual types of samples is a bit higher than the total number of \textit{Removed noisy samples} shown in the last row.
After removing noise, 421,373 Python and 344,624 Java samples are left, as shown in the \textit{Noise Free} row in Table~\ref{tab:csn}. 
We then calculate the string lengths of the docstrings for these samples. 
We assume that if a docstring is too short, it may not be informative enough for code generation. On the other hand, if a docstring is too long, it may be confusing and difficult to understand due to imprecise descriptions or noisy information. Therefore, we only select the samples whose docstring lengths are in the interval of [10\% quantile, 
% (33 for Python and 38 for Java)
90\% quantile].
% (614 for Python and 372 for Java)
Finally, 336,668 Python and 275,467 Java $<$\textit{docstring, human-written code}$>$ pairs are collected to build the final dataset, as shown in the \textit{Final} row in Table~\ref{tab:csn}, where the docstrings will be used for code generation with ChatGPT.

\begin{table}[!t]
  \centering
  \caption{Status of code generation results by ChatGPT.}
  \resizebox{0.38\linewidth}{!}
  {
    \begin{tabular}{lrr}
    \toprule
    \textbf{Status} & \multicolumn{1}{r}{\textbf{Python}} & \multicolumn{1}{r}{\textbf{Java}} \\
    \midrule
    \rowcolor{lightgray} Success & 288,508 & 222,335 \\
    Empty/Incomplete & 42,529 & 46,676 \\
    Fail  & 5,631 & 6,456 \\
    \midrule
    \textit{Number of responses} & 336,668 & 275,467 \\
    \bottomrule
    \end{tabular}%
  }
  \label{tab:success-empty-fail}%
\end{table}%

\subsubsection{Collecting ChatGPT-generated Code} \label{sec:gpt-code-collection}
We feed ChatGPT with prompts built with the docstrings from the 336,668 Python and 275,467 Java $<$\textit{docstring, human-written code}$>$ pairs to generate corresponding code. The prompt is designed with the following format:

\addvspace{0.5em}
\promptbox{
\textit{I want you to act as a software developer. I will provide you with some requirements about a function, and it will be your job to implement the function in \{language\}. Do not write explanations, just reply with the code. My request is: implement a function according to the following requirements: the function description is \{docstring\}}
}
    % \begin{tcolorbox}[boxrule=0pt,left=1pt,right=1pt,top=1pt,bottom=1pt]%
    % \end{tcolorbox}

\noindent where \{language\} is replaced by ``Python'' or ``Java'', and \{docstring\} is a placeholder for the raw docstring text.
{We collected the raw generated answers from ChatGPT through OpenAI API (``gpt-3.5-turbo'' model with default settings) in April 2023.}
We use regular expressions to remove the backquotes of code blocks and declarative sentences like \textit{``Here's the implementation of the function''}.
We also observe that ChatGPT generates empty or incomplete functions and even fails to generate any code for some cases due to insufficient information.
Eventually, 288,508 Python and 222,335 Java code snippets are successfully generated by ChatGPT.
Table~\ref{tab:success-empty-fail} lists the corresponding statistics of code generation results by ChatGPT. 
Only the successfully generated code snippets are selected to construct the \datasetname dataset.

We curate the \datasetname dataset by pairing the successfully generated code by ChatGPT with the corresponding human-written function from CSN with the same docstring.
Eventually, the \datasetname dataset contains 288K Python and 222K Java pairs of $<$\textit{human-written code, ChatGPT-generated code}$>$ pairs.
Following~\cite{husain2019codesearchnet}, we split the dataset into train, validation, and test sets with a proportion of 8:1:1 for our evaluation.

\subsection{Evaluating Baselines}

\label{baselines}
Existing AIGC detectors can be divided into commercial ones~\cite{zerogpt,chatgptzero,copyleaks,writer,aitextclassifier} and open-source ones~\cite{gpt2detector,mitchell2023detectgpt, robertaqa}. These detectors are mainly designed to detect AI-generated natural language text, such as articles and essays, and their performance in detecting AI-generated code has not been comprehensively evaluated.
To the best of our knowledge, \toolname is the first detector specially designed for detecting AI-generated program code.
To evaluate the performance of \toolname, we compare it against six existing representative AIGC detectors, namely two commercial detectors (\ie, Writer~\cite{writer} and ZeroGPT~\cite{zerogpt}),
one zero-shot detector (\ie, DetectGPT~\cite{mitchell2023detectgpt}),
and three RoBERTa-based detectors (\ie, RoBERTa-single~\cite{robertaqa}, RoBERTa-QA~\cite{robertasingle} and GPT2-Detector~\cite{huggingfacegpt2detector}). 
We give a brief introduction for the selected baselines as follows:

\begin{itemize}[leftmargin=*]
    \item \textbf{Writer}~\cite{writer}. Writer is an AI content detector that works by checking if the input text is likely to follow the same pattern of words that a LLM would produce.
    % its implementation details are not revealed.
    % the limit to input text length is 1,500 characters at a time.
    \item \textbf{ZeroGPT}~\cite{zerogpt}. ZeroGPT is a free AI content detector based on its  DeepAnalyse™ Technology. The ZeroGPT team claims that the tool reaches an accuracy rate of text detection up to 98\% in their evaluation.
    \item \textbf{DetectGPT}~\cite{mitchell2023detectgpt}.
    DetectGPT is a state-of-the-art zero-shot AIGC detection method~\cite{mitchell2023detectgpt}.
    DetectGPT compares the log probabilities of the original input text and the perturbed text obtained using another pre-trained model computed under the targeted model to decide whether the text is AI-generated.    
    \item \textbf{RoBERTa-single}~\cite{robertaqa}. Guo \etal ~\cite{guo2023hc3} build the RoBERTa-single model by fine-tuning RoBERTa~\cite{liu2019roberta} on the Human ChatGPT Comparison Corpus (HC3) dataset~\cite{guo2023hc3} which contains nearly 40K questions and corresponding answers from human experts and ChatGPT. Specifically, RoBERTa-single is trained on the text of answers from the HC3 dataset.
    \item \textbf{RoBERTa-QA}~\cite{robertasingle}. This RoBERTa-based model, also proposed by Guo \etal ~\cite{guo2023hc3}, is trained on question-answer pairs from the HC3 dataset. Guo \etal ~\cite{guo2023hc3} conclude that RoBERTa-QA is generally more effective than RoBERTa-single based on their evaluations. Since it is unknown whether this conclusion still holds in the LLM-generated code detection scenario, we include both models for comparison.
    \item \textbf{GPT2-Detector}~\cite{huggingfacegpt2detector}. 
    GPT2-Detector is the OpenAI GPT-2 Output Detector model, obtained by fine-tuning a RoBERTa base model with the outputs of the 1.5B-parameter GPT-2 model~\cite{gpt2detector}. According to the OpenAI Report~\cite{solaiman2019release}, this detector reaches an accuracy rate of about 95\% in detecting 1.5B-parameter GPT-2-generated text.
\end{itemize}

Note that both GPT2-Detector and AI Text Classifier~\cite{aitextclassifier} are proposed by OpenAI.
However, we do not consider AI Text Classifier as our baseline since it is currently shut down, as stated in its official blog~\cite{openaiblog}. 
% Besides, this tool requires a minimum input of 1,000 characters, making it unsuitable for detecting shorter code snippets present in our datasets.
% blog: As of July 20, 2023, the AI classifier is no longer available due to its low rate of accuracy.

% GPTSniffer is now accepted by Journal of Systems and Software
In addition, during the paper revision period, we noticed two concurrent works~\cite{yang2023zero, NGUYEN2024112059} released on arXiv that also focus on LLM-generated code detection. For the completeness of our experimental evaluation, we have included the methods proposed in these papers as baselines.
\begin{itemize}[leftmargin=*]
    \item \textbf{DetectGPT4Code}~\cite{yang2023zero}. DetectGPT4Code is the first zero-shot LLM-generated code detection method~\cite{yang2023zero}. DetectGPT4Code modifies DetectGPT by utilizing a small proxy model to estimate the probability of the rightmost tokens.
    \item \textbf{GPTSniffer}~\cite{NGUYEN2024112059}. Nguyen et al.~\cite{NGUYEN2024112059} propose a CodeBERT-based classifier named GPTSniffer to detect AI-written code and conduct an empirical study with GPTSniffer to assess the impact of different preprocessing settings on classification performance.
\end{itemize}

\subsection{Implementation Details}
\label{implement_details}
For commercial detectors, Writer and ZeroGPT, the results are directly collected from their official web apps by feeding with the data in the \datasetname test set. Since Writer requires the input length to be less than 1,500 characters, we truncate the input code strings to satisfy its requirement.

For zero-shot detectors, DetectGPT and DetectGPT4Code, we follow their publicly available implementation on GitHub~\cite{pytorchdetectgpt, detectgpt4codegithub}. For DetectGPT, we follow~\cite{pytorchdetectgpt} to set the average log ratio classification threshold as 0.7 and use T5~\cite{t5large} and GPT2-medium~\cite{gpt2medium} as the perturbation model and source model, respectively.
For DetectGPT4Code, we follow ~\cite{detectgpt4codegithub} to set the number of perturbations $N$ as 20 and ratio $\gamma$ as 0.9, and report performance metrics at 10\% FPR. DetectGPT4Code uses InCoder-6B~\cite{InCoder6B} as the perturbation model and uses PolyCoder-160M~\cite{PolyCoder160M} as the surrogate model.

For training-based methods, we obtain the RoBERTa-based detectors (i.e., RoBERTa-single, RoBERTa-QA, and GPT2-Detector) from Hugging Face \textit{transformers}~\cite{wolf2020transformers}. 
We follow the publicly available implementation on GitHub for GPTSniffer~\cite{gptsniffergithub}.
For \toolname, we utilize unixcoder-base-nine~\cite{unixcoderhuggingface} as the basic semantic encoder. During fine-tuning, we follow Guo et al.~\cite{guo2022unixcoder} to set the learning rate as 2e-5 and the batch size as 8. We use the Adam optimizer to fine-tune the model and set the maximum number of epochs as 20 to ensure sufficient convergence and perform early stopping on the validation set to prevent over-fitting.
The models with optimal performance on the validation set are used for the evaluations.
The classification threshold is set as 0.5 for all training-based methods.

All the models are implemented in Python using the \textit{PyTorch} framework. Experiments are performed in Ubuntu 20.04 with an Nvidia GeForce RTX 3090 GPU driven by CUDA 12.0.

\subsection{Experimental Design}

In this section, we introduce the research questions and the corresponding experimental setup of this study.

1) {\bf\em RQ-1: To what extent can ChatGPT-generated code be manually identified by developers?}
We conduct a human study with experienced developers to investigate whether they can effectively distinguish ChatGPT-generated code from human-written code with their professional knowledge.

\noindent
\textbf{Experimental setup:} For Java and Python, we randomly select 10 ChatGPT-generated and 10 human-written functions from the \datasetname dataset, respectively. We remove all the code comments, considering that they might interfere with participants' decisions.
The survey is conducted on an online survey platform~\cite{wjx}. We design separate questionnaires for Java and Python. Each questionnaire contains 20 true or false questions and one optional open-ended question. For each true or false question, participants are asked to determine whether the function is written by ``ChatGPT'' or ``Human''. For the open-ended question, participants are asked to briefly explain the reasoning behind their judgments. 
The participants in our study are software developers with three to five years of programming experience.

2) {\bf\em RQ-2: What are the characteristics of ChatGPT-generated codes?}
We analyze the characteristics of ChatGPT-generated code from qualitative and quantitative aspects to overview the differences between ChatGPT-generated and human-written codes.

\noindent
\textbf{Experimental setup:} 
To facilitate our qualitative analysis, we analyze 200 random docstring-response pairs, 100 for Python and 100 for Java (95\% confidence level +/- 10\%), to determine how ChatGPT responds to our designed prompts for code generation.
Furthermore, we examine 100 random ChatGPT-generated and human-written code pairs of each programming language (95\% confidence level +/- 10\%) to explore the characteristic differences between ChatGPT-generated and human-written code.

% pair/pair_java_with_features.jsonl.gz 221844 
% pair/pair_py_with_features.jsonl.gz 288461
For quantitative analysis, we compute the following five metrics across the entire \datasetname dataset (i.e., 288K Python and 222K Java human-written and ChatGPT-generated code pairs) to mine more features:

\begin{itemize} 
    \item \textbf{Number of Lines Of Code (NLOC).} It is calculated by the total lines of code after excluding blank lines and comments.
    NLOC is a common metric for quantifying the size of a software program but is an insufficient proxy for measuring software complexity~\cite{tashtoush2014correlation}.
    \item \textbf{Cyclomatic Complexity Number (CCN).} The CCN metric, developed by McCabe~\cite{mccabe1976complexity}, is one of the most popular metrics to measure the complexity of a software program. CCN is calculated by the number of linearly independent paths through the source code. A larger CCN indicates that the code structure is more complex.
    \item \textbf{Token Count (TC).} It is calculated by the total number of tokens (e.g., keywords, identifiers, constants, operators) contained in the source code.
    \item \textbf{Length of the Function Name (LFN).} It is calculated by the string length of the function name. For example, the function name of \code{def add(self, r)} is \code{add}, whose LFN is 3.
    \item \textbf{Unique tokens} represents the total number of unique tokens in the corpus.
\end{itemize}

% \r{
The \textit{NLOC} and \textit{TC} metrics quantify code size, while the \textit{CCN} metric evaluates code complexity. Additionally, \textit{LFN} and \textit{unique tokens} provide insights into the distinctions of naming conventions and vocabulary. We calculate these metrics with a code complexity analyzer named \textit{lizard}~\cite{lizard}, and a fast open-source tokenizer named \textit{tiktoken}~\cite{tiktoken}.

3) {\bf\em RQ-3: Can \toolname differentiate ChatGPT-generated code from human-written code effectively?}

We compare the performance of our ChatGPT-generated code detection model \toolname with the selected commercial and open-source baselines on the \datasetname test set.

\noindent
\label{param_baselines}
\textbf{Experimental setup:} 
We follow the default settings of the baseline detectors, as introduced in Section~\ref{implement_details}.
In particular, since DetectGPT and the RoBERTa-based detectors are designed to detect natural language text instead of source code, we additionally implement their improved versions to ensure a fair comparison. 
Specifically, for DetectGPT, we replace the perturbation model T5 with code pre-trained models: InCoder-6B~\cite{InCoder6B} and CodeLlama-7B~\cite{CodeLlama7b}, represented as \textit{DetectGPT* (w/ InCoder)} and \textit{DetectGPT* (w/ CodeLlama)}. We fine-tune the RoBERTa-based detectors on the \datasetname training set, represented as \textit{RoBERTa-single*}, \textit{RoBERTa-QA*}, and \textit{GPT2-Detector*}. 
GPTSniffer is also fine-tuned on the \datasetname training set for a fair comparison. 
We evaluate both the default baseline detectors and our improved versions on the \datasetname test set.

4) {\bf\em Evaluation metrics:}
\label{sec:eval_metrics}
To evaluate the effectiveness of existing AIGC detectors and \toolname on detecting ChatGPT-generated code, we adopt the following evaluation metrics: Accuracy, Precision, Recall, F1-score, and AUC, which are widely used in the literature~\cite{ni2022best,ni2020jitjs,fu2022vulrepair}.

For each function, there are four possible detection results:
it can be detected as ChatGPT-generated if it is exactly generated by ChatGPT (true positive, TP); 
it can be detected as ChatGPT-generated while it is human-written (false positive, FP); 
it can be detected as human-written while it is ChatGPT-generated (false negative, FN); 
or it can be detected as human-written when it is exactly human-written (true negative, TN).
Therefore, based on the four possible results, evaluation metrics can be defined as follows:

\textbf{Accuracy} evaluates the performance of how many functions can be correctly classified.
It is calculated as $\frac{TP+ TN}{TP+FP+TN +FN}$.

\textbf{Precision} is the fraction of ChatGPT-generated functions among the detected positive instances, which can be calculated as $\frac{TP}{TP+FP}$.

\textbf{Recall} measures how many ChatGPT-generated functions can be correctly detected, which is defined as $\frac{TP}{TP+FN}$.

\textbf{F1-score} is a harmonic mean of $Precision$ and $Recall$, which can be calculated as $\frac{2 \times P \times R}{P + R}.$

\textbf{AUC} is the area under the receiver operating characteristic (ROC) curve, which is a 2D illustration of true positive rate (TPR) on the y-axis versus false positive rate (FPR) on the x-axis. 
ROC curve can be obtained by varying the classification threshold over all possible values.
The AUC score ranges from 0 to 1, and a well-performing classifier provides a value close to 1.

%% file: sections/experiment_results.tex
\section{Experiment Results}
\label{sec:results}
\subsection{\bf{[RQ-1]: Ineffective Manual Identification.}} 
\label{human_study_results}
By April 27, 2023, we have received \textbf{21} and \textbf{23} valid responses to Python and Java questionnaires, respectively, after distributing 30 questionnaires for each programming language.
The average accuracy rate of the Python group is \textbf{50\%}, and the average accuracy rate of the Java group is \textbf{49.75\%}, which is close to the random guess.
The results suggest that distinguishing ChatGPT-generated from human-written code only based on developers' own knowledge is a big challenge for them, even if they are familiar with the programming language. 
Besides, we observe that participants falsely classified 58.86\% ChatGPT-generated code as human-written while falsely classified 41.59\% human-written code as ChatGPT-generated, which reveals that participants are prone to consider the code as human-written.

With 31 responses from participants for the open-ended question, the reasons behind their decisions can be summarized into three main categories: \textbf{coding style} (18 mentions), \textbf{code logic} (15 mentions), and \textbf{intuition} (5 mentions).
For instance, a participant mentioned: ``{\em The variable names generated by ChatGPT are relatively long, while human developers favor concise variable names}.'' while another reported: ``{\em I think variables generated by ChatGPT often have repeated words with parameters and the function name, while other types of variable names, such as variables named according to common sense, may occur in functions written by human}.''
The other participant made judgments with coding style and replied: ``{\em ChatGPT strictly follows the variable naming conventions, and typically does not write a very long line of code}.''
From a code logic perspective, several participants claimed that ``{\em ChatGPT-generated functions perform relatively simpler operations}''.
Moreover, a few participants failed to find a baseline for their judgments. For instance, a participant mentioned that ``{\em It is basically hard to tell. I just followed my intuition and experience of using ChatGPT}''.

\addvspace{0.2em}
\intuition{
\textbf{Answer to RQ-1}: The human participants cannot effectively differentiate ChatGPT-generated code from human-written code, achieving only a 50\% accuracy, which is close to random selection. Most participants rely on factors such as coding style, code logic, and even intuition to make their identifications. 
These findings underscore the necessity of employing tools, such as deep learning models, to enhance the accuracy of detecting LLM-generated code.
}

\subsection{\bf{[RQ-2]: Qualitative and Quantitative Characteristics.}}
\label{sec:rq1}
\subsubsection{Qualitative Analysis} 
In this section, we analyze the qualitative characteristics of ChatGPT-generated code based on manual observation after analyzing 200 docstring-response pairs and 200 pairs of ChatGPT-generated and human-written code.

\begin{figure*}[t]
\centerline{
    \includegraphics[width=0.9\linewidth]{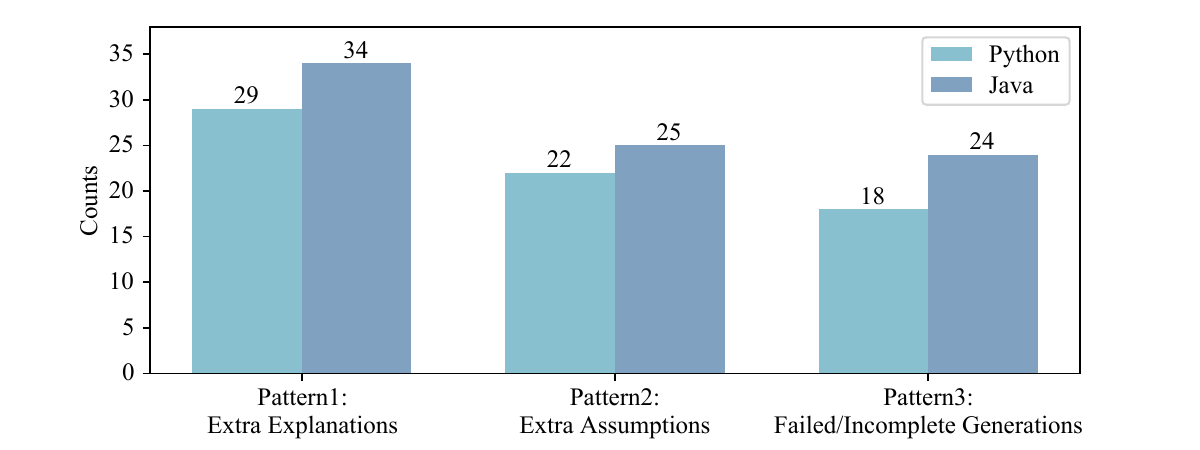}
    }
    \caption{The numerical counts of cases corresponding to each observation.}
    \label{fig:rq2_response}
    \Description{This bar chart compares the counts of three patterns between two programming languages, Python and Java. For Pattern 1: Extra Explanations, Python has a count of 29, while Java has a higher count of 34. 
    For Pattern 2: Extra Assumptions, Python shows a count of 22, and Java has a higher count of 25.
    For Pattern 3: Failed/Incomplete Generations, Python has a count of 18, while Java has a higher count of 24.}
\end{figure*}

1) \textit{Patterns of raw ChatGPT responses:} After examining 100 Python and 100 Java docstring-response pairs, we observe that:
\ding{182} \textbf{ChatGPT tends to offer explanations about what the generated code does and how to use it.} Although our prompts explicitly mention with ``{\em Do not write explanations}'', ChatGPT still often writes declarative sentences to describe the logic of the generated function and possible dependencies required to execute the code. 
For some cases, ChatGPT provides additional code snippets with examples explaining how to use the generated function.
\ding{183} \textbf{ChatGPT may assume additional conditions that are not provided in the docstring of prompts.} In an effort to generate the function, ChatGPT would guess the possible implementing details of the function (\eg, ``{\em I assumed that the function should return 1 on success...}'') and fabricate global variables or custom classes that may not actually exist (\eg, ``{\em This assumes that CommandLineException is a custom exception class that you have defined}'').
\ding{184} \textbf{ChatGPT refuses to generate code when it thinks the required information is insufficient.} 
In the cases where requirements are vague, information about dependent libraries is lacking, or the implementation involves specific technologies beyond its knowledge, ChatGPT may either refuse to generate code or respond by only providing the method signature.
The numerical counts of cases corresponding to the three patterns mentioned above are reported in Figure~\ref{fig:rq2_response}.

\label{finding:skeleton}
\textit{Characteristics of ChatGPT-generated code:} After examining 100 Python and 100 Java pairs of ChatGPT-generated and human-written code, we observe that:
\ding{182} \textbf{ChatGPT tends to extract keywords from the docstring to generate function names.}
{In almost all cases (around 95\% in our observations),} ChatGPT-generated function names are sourced from the provided docstring with the \code{under\_score} form for Python code and the \code{camelCase} form for Java code, while developers could use words that are not included in the docstring with various naming conventions.
For instance, given a docstring of a Python function ``{\em Add a request, response pair to this cassette}'', the ChatGPT-generated function name is \code{add\_request\_response\_pair}, while the human-written one is \code{append}.
\ding{183} \textbf{ChatGPT avoids using meaningless words or characters to name variables.} {In almost all cases (around 98 \% in our observation),} ChatGPT tends to use informative and meaningful words to name variables (\eg, ``total\_samples'', ``numCharsRead''). Besides, ChatGPT uses common abbreviations (\eg, abbreviate ``object'' as ``obj'', ``function'' as ``func'') and uses the acronym of the class name to name its instance (\eg, ``FileInputStream fis'', ``StringBuilder sb''). In contrast, developers are prone to use informal abbreviations in variable names (\eg, abbreviate ``object'' as ``o'', ``submission\_id'' as ``s\_id'') for more cases.
\ding{184} \textbf{ChatGPT provides a skeleton of the function when specific implementation details are unknown.}
{In certain cases (around 14\% in our observations),} when implementation details are not explicitly included in the docstring, ChatGPT will generate a template function with comments like ``\code{\# replace with your implementation}'' to suggest
where the custom implementations should be filled in or which statements should be modified according to actual requirements.

\subsubsection{Quantitative Analysis}

\begin{table}[tbp]
  \centering
  \caption{The code metrics of ChatGPT-generated and Human-written code.}
  \tabcolsep=0.1cm
  \resizebox{0.75\linewidth}{!}
  {
    \begin{tabular}{ccrrrrr}
    \toprule
    \textbf{Language} & \textbf{Group} & \textbf{Avg. NLOC} & \textbf{Avg. CCN} & \textbf{Avg. TC} &\textbf{ Avg. LFN} & \textbf{Unique tokens} \\
    \midrule
    \multirow{2}[2]{*}{Python} & ChatGPT & 8.07  & 2.52  & 57.20  & \textbf{17.56} & 63,342  \\
          & Human & \textbf{13.71} & \textbf{4.02} & \textbf{104.22} & 13.64  & \textbf{73,912} \\
    \midrule
    \multirow{2}[2]{*}{Java} & ChatGPT & 10.77  & 2.51  & 78.49  & \textbf{18.35} & 53,915  \\
          & Human & \textbf{14.66} & \textbf{3.72} & \textbf{106.99} & 13.59  & \textbf{57,870} \\
    \bottomrule
    \end{tabular}%
  }
  \label{tab:code_features}%
\end{table}%

% ========================
\label{sec:statistical}
In this section, we analyze the quantitative features of ChatGPT-generated and human-written code in our \datasetname dataset, of which related code metrics are shown in Table~\ref{tab:code_features}.
Based on the results and our observations, we can draw the following conclusions: 
\ding{182} \textbf{ChatGPT-generated functions are relatively simpler than human-written ones.} The values of \textit{Avg. NLOC}, \textit{Avg. CCN} and \textit{Avg. TC} of human-written code are higher than those of ChatGPT-generated code, which suggests that human-written functions normally present more complex logic and structures than ChatGPT-generated ones.
This is consistent with the observation presented in the previous paragraph
that human-written functions implement every detailed functional requirement, while ChatGPT often provides a sample implementation structure where specific implementations are not generated.
\ding{183} \textbf{ChatGPT-generated functions have longer function names than human-written ones.} As discussed in the previous paragraph, ChatGPT-generated function names often consist of both verbs and nouns, while human developers sometimes use a single noun or verb to name functions, or even use abbreviations to make function names concise.
\ding{184} \textbf{ChatGPT-generated codes have a relatively smaller vocabulary size than human-written ones.} We observe that the tokens in ChatGPT-generated code are often sourced from the provided docstring, standard and popular Python or Java libraries (e.g., \textit{numpy}, \textit{re}, \textit{sys}), or commonly used variable names (e.g., ``name'', ``root'', ``key''), while human developers apply custom libraries or classes for code implementation, and use more diverse words to name variables. 

\addvspace{0.2em}
\intuition{
\textbf{Answer to RQ-2}: 
If requirements are insufficiently detailed in prompts, ChatGPT may fabricate information or generate an incomplete implementation.
ChatGPT is prone to use keywords from prompts to name the functions and avoids using meaningless words to name variables. 
ChatGPT-generated functions are less complex, with a smaller vocabulary size but more verbose function names.
}

\subsection{\bf{[RQ-3]: Effectiveness of \toolname.}}

\subsubsection{Comparing with the Selected Baselines} 
Table~\ref{tab:with_comment} shows the results of differentiating ChatGPT-generated code from human-written code with the selected baselines and \toolname.
According to the results, the accuracy of the commercial baselines, Writer and ZeroGPT, is no higher than 0.537 and 0.552 on Python and Java test sets, respectively. Writer and ZeroGPT tend to classify code snippets as human-written, leading to a low recall rate.
Regarding the zero-shot detectors, the original DetectGPT performs poorly, with an accuracy of less than 0.5. We observe that DetectGPT tends to label code snippets as ChatGPT-generated, which explains its extremely high recall.
Our improved versions of DetectGPT show some improvement, with DetectGPT* (w/ InCoder) achieving the highest accuracy and AUC on the Python test set among all the zero-shot methods. 
% Overall, DetectGPT4Code is the best zero-shot detector. 
Regarding Accuracy and AUC, DetectGPT4Code outperforms DetectGPT, especially on the Java datasets.
However, despite being designed to detect AI-generated code, its accuracy is low on our \datasetname test set. We believe that carefully adjusting hyperparameters such as the number of perturbations, truncation ratio, classification threshold, and number of masked lines could further improve its performance, but this process would be very time-consuming.
In summary, we can conclude that \textbf{detecting ChatGPT-generated code is a big challenge for the existing commercial and zero-shot AIGC detectors.}

\begin{table*}[t]
  \centering
  \caption{Results of different methods on the \datasetname test set. The best results are highlighted in bold.}
  \resizebox{\linewidth}{!}{
    % Table generated by Excel2LaTeX from sheet '【major】'
    \begin{tabular}{cl|ccccc|ccccc}
    \toprule
    \multicolumn{2}{c|}{\multirow{2}[4]{*}{\textbf{Method}}} & \multicolumn{5}{c|}{\textbf{Python}}  & \multicolumn{5}{c}{\textbf{Java}} \\
    \cmidrule{3-12}\multicolumn{2}{c|}{} & \textbf{Accuracy} & \textbf{Recall} & \textbf{Precision} & \textbf{F1-score} & \textbf{AUC} & \textbf{Accuracy} & \textbf{Recall} & \textbf{Precision} & \textbf{F1-score} & \textbf{AUC} \\
    \midrule
    \multirow{2}[2]{*}{Commercial} & Writer & 0.537  & 0.122  & 0.721  & 0.209  & 0.667  & 0.455  & 0.092  & 0.336  & 0.144  & 0.378  \\
          & ZeroGPT & 0.536  & 0.086  & 0.851  & 0.157  & 0.549  & 0.552  & 0.217  & 0.658  & 0.326  & 0.586  \\
    \midrule
    \multirow{4}[2]{*}{Zero-shot} & DetectGPT & 0.499  & 0.985  & 0.499  & 0.662  & 0.546  & 0.477  & 0.905  & 0.487  & 0.635  & 0.471  \\
          & {- DetectGPT* (w/ InCoder)} & {0.565 } & {0.171 } & {0.873 } & {0.286 } & {0.560 } & {0.487 } & {0.099 } & {0.456 } & {0.163 } & {0.469 } \\
          & {- DetectGPT* (w/ CodeLlama)} & {0.517 } & {0.069 } & {0.789 } & {0.127 } & {0.423 } & {0.480 } & {0.027 } & {0.315 } & {0.050 } & {0.416 } \\
          & {DetectGPT4Code} & {0.565 } & {0.242 } & {0.715 } & {0.362 } & {0.505 } & {0.509 } & {0.123 } & {0.555 } & {0.201 } & {0.500 } \\
    \midrule
    \multicolumn{1}{c}{\multirow{3}{2cm}{\centering Default \\ Training-based}} & RoBERTa-single & 0.492  & 0.022  & 0.367  & 0.042  & 0.428  & 0.500  & 0.000  & 0.286  & 0.001  & 0.479  \\
          & RoBERTa-QA & 0.501  & 0.003  & 0.594  & 0.006  & 0.746  & 0.500  & 0.000  & 0.313  & 0.001  & 0.470  \\
          & GPT2Detector & 0.606  & 0.415  & 0.671  & 0.513  & 0.672  & 0.500  & 0.371  & 0.500  & 0.426  & 0.491  \\
    \midrule
    \multicolumn{1}{c}{\multirow{5}{2cm}{\centering Fine-tuned \\ Training-based}}
          & - RoBERTa-single* & 0.981  & 0.983  & 0.980  & 0.981  & 0.998  & 0.937  & 0.949  & 0.927  & 0.938  & 0.985  \\
          & - RoBERTa-QA* & 0.984  & 0.989  & 0.980  & 0.984  & \textbf{0.999 } & 0.946  & 0.957  & 0.936  & 0.946  & 0.988  \\
          & - GPT2-Detector* & 0.986  & 0.989  & 0.983  & 0.986  & \textbf{0.999 } & 0.946  & 0.961  & 0.933  & 0.947  & 0.989  \\
          & {GPTSniffer} & {0.985 } & {\textbf{0.991 }} & {0.979 } & {0.985 } & {\textbf{0.999 }} & {0.950 } & {0.936 } & {\textbf{0.963 }} & {0.949 } & {0.990 } \\ \rowcolor{lightgray}
          & \toolname & \textbf{0.992 } & \textbf{0.991 } & \textbf{0.993 } & \textbf{0.992 } & \textbf{0.999 } & \textbf{0.967 } & \textbf{0.976 } & 0.959  & \textbf{0.968 } & \textbf{0.995 } \\
    \bottomrule
    \end{tabular}%
  }
  \label{tab:with_comment}%
\end{table*}%

\begin{table*}[tbp]
  \centering
  \caption{Results of different methods on the \datasetname-no-comments test set. The best results are highlighted in bold. The value inside ``()'' indicates the percentage by which the metric increases or decreases compared to the result on the original \datasetname test set with code comments.}
  \resizebox{\linewidth}{!}{
    % Table generated by Excel2LaTeX from sheet '【major】'
    \begin{tabular}{cl|lllll|lllll}
    \toprule
    \multicolumn{2}{c|}{\multirow{2}[4]{*}{\textbf{Method}}} & \multicolumn{5}{c|}{\textbf{Python-no-comments}} & \multicolumn{5}{c}{\textbf{Java-no-comments}} \\
    \cmidrule{3-12}\multicolumn{2}{c|}{} & \multicolumn{1}{c}{\textbf{Accuracy}} & \multicolumn{1}{c}{\textbf{Recall}} & \multicolumn{1}{c}{\textbf{Precision}} & \multicolumn{1}{c}{\textbf{F1-score}} & \multicolumn{1}{c|}{\textbf{AUC}} & \multicolumn{1}{c}{\textbf{Accuracy}} & \multicolumn{1}{c}{\textbf{Recall}} & \multicolumn{1}{c}{\textbf{Precision}} & \multicolumn{1}{c}{\textbf{F1-score}} & \multicolumn{1}{c}{\textbf{AUC}} \\
    \midrule
    \multirow{2}[2]{*}{Commercial} & Writer & 0.513 (-4.6\%) & 0.173 (41.7\%) & 0.541 (-25.0\%) & 0.262 (25.6\%) & 0.536 (-19.6\%) & 0.468 (2.7\%) & 0.116 (26.6\%) & 0.392 (16.5\%) & 0.179 (24.3\%) & 0.447 (18.2\%) \\
          & ZeroGPT & 0.523 (-2.3\%) & 0.108 (24.6\%) & 0.638 (-25.1\%) & 0.184 (17.4\%) & 0.523 (-4.7\%) & 0.567 (2.7\%) & 0.248 (14.4\%) & 0.684 (4.0\%) & 0.364 (11.7\%) & 0.592 (1.0\%) \\
    \midrule
    \multirow{4}[2]{*}{Zero-shot} & DetectGPT & 0.497 (-0.4\%) & \textbf{0.984} (-0.1\%) & 0.498 (-0.3\%) & 0.661 (-0.2\%) & 0.545 (-0.2\%) & 0.469 (-1.6\%) & 0.905 (0.0\%) & 0.483 (-0.7\%) & 0.630 (-0.7\%) & 0.465 (-1.3\%) \\
          & {- w/ InCoder} & {0.522 (-7.7\%)} & {0.188 (10.0\%)} & {0.581 (-33.4\%)} & {0.284 (-0.6\%)} & {0.539 (-3.7\%)} & {0.500 (2.8\%)} & {0.134 (35.8\%)} & {0.516 (13.2\%)} & {0.213 (31.1\%)} & {0.506 (7.8\%)} \\
          & {- w/ CodeLlama} & {0.503 (-2.7\%)} & {0.107 (55.6\%)} & {0.535 (-32.2\%)} & {0.179 (41.0\%)} & {0.531 (25.4\%)} & {0.504 (4.9\%)} & {0.072 (165.2\%)} & {0.558 (77.1\%)} & {0.128 (155.1\%)} & {0.491 (18.0\%)} \\
          & {DetectGPT4Code} & {0.496 (-12.3\%)} & {0.099 (-59.0\%)} & {0.503 (-29.7\%)} & {0.166 (-54.2\%)} & {0.497 (-1.7\%)} & {0.543 (6.7\%)} & {0.191 (55.2\%)} & {0.659 (18.8\%)} & {0.296 (47.0\%)} & {0.572 (14.3\%)} \\
    \midrule
    \multicolumn{1}{c}{\multirow{3}{2cm}{\centering Default \\ Training-based}} & RoBERTa-single & 0.500 (1.6\%) & 0.002 (-90.6\%) & 0.455 (23.9\%) & 0.004 (-90.2\%) & 0.524 (22.3\%) & 0.500 (0.0\%) & 0.000 (-25.0\%) & 0.231 (-19.2\%) & 0.001 (-28.6\%) & 0.487 (1.6\%) \\
          & RoBERTa-QA & 0.474 (-5.2\%) & 0.018 (535.7\%) & 0.205 (-65.5\%) & 0.033 (475.4\%) & 0.338 (-54.7\%) & 0.500 (0.0\%) & 0.001 (25.0\%) & 0.353 (12.9\%) & 0.001 (22.2\%) & 0.543 (15.5\%) \\
          & GPT2-Detector & 0.548 (-9.6\%) & 0.498 (19.9\%) & 0.553 (-17.6\%) & 0.524 (2.2\%) & 0.558 (-17.0\%) & 0.519 (3.7\%) & 0.408 (10.1\%) & 0.524 (4.8\%) & 0.459 (7.8\%) & 0.524 (6.8\%) \\
    \midrule
    \multicolumn{1}{c}{\multirow{5}{2cm}{\centering Fine-tuned \\ Training-based}}
          & - RoBERTa-single* & 0.949 (-3.3\%) & 0.961 (-2.3\%) & 0.939 (-4.2\%) & 0.950 (-3.2\%) & 0.990 (-0.8\%) & 0.932 (-0.6\%) & 0.941 (-0.8\%) & 0.924 (-0.4\%) & 0.932 (-0.6\%) & 0.981 (-0.4\%) \\
          & - RoBERTa-QA* & 0.955 (-3.0\%) & 0.970 (-1.9\%) & 0.941 (-4.0\%) & 0.955 (-3.0\%) & 0.992 (-0.7\%) & 0.938 (-0.8\%) & 0.953 (-0.4\%) & 0.925 (-1.1\%) & 0.939 (-0.8\%) & 0.986 (-0.2\%) \\
          & - GPT2-Detector* & 0.959 (-2.7\%) & 0.971 (-1.8\%) & 0.948 (-3.5\%) & 0.959 (-2.7\%) & 0.993 (-0.6\%) & 0.941 (-0.5\%) & 0.958 (-0.4\%) & 0.928 (-0.6\%) & 0.942 (-0.5\%) & 0.987 (-0.2\%) \\
          & {GPTSniffer} & {0.959 (-2.6\%)} & {0.973 (-1.8\%)} & {0.946 (-3.3\%)} & {0.960 (-2.6\%)} & {0.993 (-0.6\%)} & {0.942 (-0.9\%)} & {0.951 (1.7\%)} & {0.934 (-3.1\%)} & {0.942 (-0.7\%)} & {0.986 (-0.3\%)} \\ \rowcolor{lightgray}
          & \toolname & \textbf{0.973} (-1.9\%) & 0.975 (-1.6\%) & \textbf{0.970} (-2.3\%) & \textbf{0.973} (-1.9\%) & \textbf{0.995} (-0.4\%) & \textbf{0.964} (-0.4\%) & \textbf{0.968} (-0.8\%) & \textbf{0.959} (0.0\%) & \textbf{0.964} (-0.4\%) & \textbf{0.994} (-0.1\%) \\
    \bottomrule
    \end{tabular}%
  }
  \label{tab:result_no_comment}%
\end{table*}%

\begin{table*}[tbp]
  \centering
  \caption{Results of comparing \toolname with RoBERTa-single*, RoBERTa-QA*, GPT2-Detector*, and GPTSniffer by McNemar’s test and Odds Ratio (OR) effect size measure, where $p$-value $< 0.05$ indicates a statistically significant difference and OR $> 1$ is in favor of \toolname.}
  \resizebox{0.75\linewidth}{!}{
  {
    % Table generated by Excel2LaTeX from sheet 'final_res'
    \begin{tabular}{l|cc|cc|cc|cc}
    \toprule
    \multicolumn{1}{c|}{\multirow{2}[4]{*}{Method}} & \multicolumn{2}{c|}{Python} & \multicolumn{2}{c|}{Java} & \multicolumn{2}{c|}{Python-no-comments} & \multicolumn{2}{c}{Java-no-comments} \\
    \cmidrule{2-9}      & $p$-value & OR    & $p$-value & OR    & $p$-value & OR    & $p$-value & OR \\
    \midrule
    RoBERTa-single* & 1.66E-42 & 127   & 3.25E-81 & 36    & 4.43E-15 & 22    & 3.42E-56 & 24  \\
    RoBERTa-QA* & 5.81E-27 & 180   & 9.58E-51 & 46    & 1.85E-28 & 27    & 1.92E-31 & 32  \\
    GPT2-Detector*  & 3.49E-22 & 229   & 7.80E-50 & 48    & 2.14E-34 & 27    & 7.74E-31 & 34  \\
    GPTSniffer & 1.27E-25 & 253   & 5.78E-35 & 45    & 1.99E-33 & 36    & 4.15E-20 & 34  \\
    \bottomrule
    \end{tabular}%
  }
  }
  \label{tab:p_value}%
\end{table*}%

For training-based methods, the original RoBERTa-single, RoBERTa-QA, and GPT2Detector also show poor accuracy, with most results around 0.5. The only exception is GPT2Detector, which achieves an accuracy of 0.606 on the Python test set. 
We observe that RoBERTa-single and RoBERTa-QA almost always misclassify ChatGPT-generated code as human-written, resulting in an extremely low recall that is close to zero. 
Our fine-tuned versions of RoBERTa-based detectors achieve scores higher than 0.9 in all metrics, representing a significant improvement compared to their original versions. This indicates that \textbf{fine-tuning a pre-trained model with a large-scale dataset of human-written and LLM-generated code can be a feasible solution to detect LLM-generated code}. 
A recently released paper, \textit{independent and concurrent to our work}, explores this idea by fine-tuning CodeBERT on their dataset consisting of around 1.5K code snippets and names their method as GPTSniffer~\cite{NGUYEN2024112059}. As a comparison, we fine-tune GPTSniffer on \datasetname and report its performance. 
Overall, \toolname outperforms all selected baselines.
Since all fine-tuned training-based methods achieve competitive performance, we follow Nguyen et al.~\cite{NGUYEN2024112059} to calculate $p$-values using McNemar’s test~\cite{mcnemar1947note} with Holm’s correction, complemented by the Odds Ratio (OR) effect size measure, for further comparisons. As shown in Table~\ref{tab:p_value}, McNemar’s test indicates statistically significant differences with ORs in favor of \toolname. Therefore, we can conclude that \textbf{\toolname is more effective than the selected baselines in distinguishing between ChatGPT-generated code and human-written code}.

As discussed in Section~\ref{finding:skeleton}, ChatGPT-generated code may contain obvious indicating words in code comments (\eg, \textit{``replace with your implementation''}), which may impact the detection results.
To investigate the influence of code comments on the detectors, we remove all comments from the code snippets to create a \textit{no-comments} variant of the \datasetname dataset and re-evaluate all the methods. 
According to the results shown in Table~\ref{tab:result_no_comment}, the conclusion that fine-tuned training-based methods perform the best still holds on the \datasetname-no-comments test set. We further compare \toolname with other fine-tuned training-based methods using McNemar’s test, and results in Table~\ref{tab:p_value} show statistically significant differences with ORs in favor of \toolname.
Regarding performance metric variations, we observe that commercial detectors are significantly impacted by code comments. For example, on the \textit{Python-no-comments} test set, the AUC scores of Writer and ZeroGPT drop by 19.6\% and 4.7\%, respectively.
As for zero-shot methods, DetectGPT is slightly influenced by code comments, where the metrics drop by no more than 1.6\%, while the performances of DetectGPT* (w/ InCoder), DetectGPT* (w/ CodeLlama), and DetectGPT4Code are less stable.
As for training-based methods, the original RoBERTa-single, RoBERTa-QA, and GPT2-Detector detectors are significantly impacted by code comments, while the performances of the fine-tuned ones seem more stable. 
In comparison with other fine-tuned training-based methods, \textbf{\toolname is less susceptible to the influence of code comments}, with only a 1.9\% drop in F1-score and 0.4\% drop in AUC on the \textit{Python-no-comments} test set, and only 0.4\% drop in F1-score and 0.1\% drop in AUC on the \textit{Java-no-comments} test set.
The better robustness of \toolname may be owing to its basic semantic encoder, namely UniXcoder, which takes code comments and AST as multi-modal contents, while the baselines treat both code and code comments equally as text sequence.

% ------ [2024/08/04] revision 2 gpt4
\subsubsection{Comparing with GPT-4}
{
Considering GPT-4's heightened text comprehension capabilities and inspired by Bhattacharjee et al.'s work that uses GPT-4 as a detector for AI-generated articles~\cite{bhattacharjee2024fighting}, we also investigate GPT-4's performance in detecting ChatGPT-generated code. 
Given the extensive scale of \datasetname and the significant costs of interacting with GPT-4, we follow previous work~\cite{croft2023data} to statistically sample a subset of cases from the \datasetname test set with a 95\% confidence level and evaluate GPT-4's performance on the sampled code snippets, which consist of 50 human-written and 50 ChatGPT-generated code snippets for each programming language. 
We use the following prompt to evaluate the zero-shot performance of GPT-4 in detecting LLM-generated code:
}

\promptbox{
Task: Identify whether the given code snippet is generated by AI or is human-written. Output only the binary classification result: "True" if it is AI-generated, and "False" if it is human-written. 
\\Code snippet to identify: <code>
}

\begin{table*}[tbp]
  \centering
  \caption{Results of comparing GPT-4, GPTSniffer, and \toolname for detecting ChatGPT-generated code on the sampled Java and Python subsets.}
  \resizebox{0.7\linewidth}{!}{
  {
    % Table generated by Excel2LaTeX from sheet 'result_metrics'
    \begin{tabular}{lllllll}
    \toprule
    \textbf{Language} & \textbf{Method} & \textbf{Accuracy} & \textbf{Recall} & \textbf{Precision} & \textbf{F1-score} & \textbf{AUC} \\
    \midrule
    \multirow{3}[2]{*}{Python} & gpt-4-0613 & 0.520  & 0.040  & \textbf{1.000 } & 0.077  & 0.382  \\
          & GPTSniffer & 0.970  & \textbf{0.980 } & 0.961  & 0.970  & \textbf{0.999 } \\
          & \toolname & \textbf{0.980 } & \textbf{0.980 } & 0.980  & \textbf{0.980 } & \textbf{0.999 } \\
    \midrule
    \multirow{3}[2]{*}{Java} & gpt-4-0613 & 0.500  & 0.000  & 0.000  & 0.000  & 0.486  \\
          & GPTSniffer & 0.920  & \textbf{0.920 } & 0.920  & 0.920  & 0.975  \\
          & \toolname & \textbf{0.940 } & 0.900  & \textbf{0.978 } & \textbf{0.938 } & \textbf{0.979 } \\
    \bottomrule
    \end{tabular}%
   }
  }
  \label{tab:rq3_gpt4}%
\end{table*}%

\begin{figure*}[tbp]
\centerline{
    \includegraphics[width=0.6\linewidth]{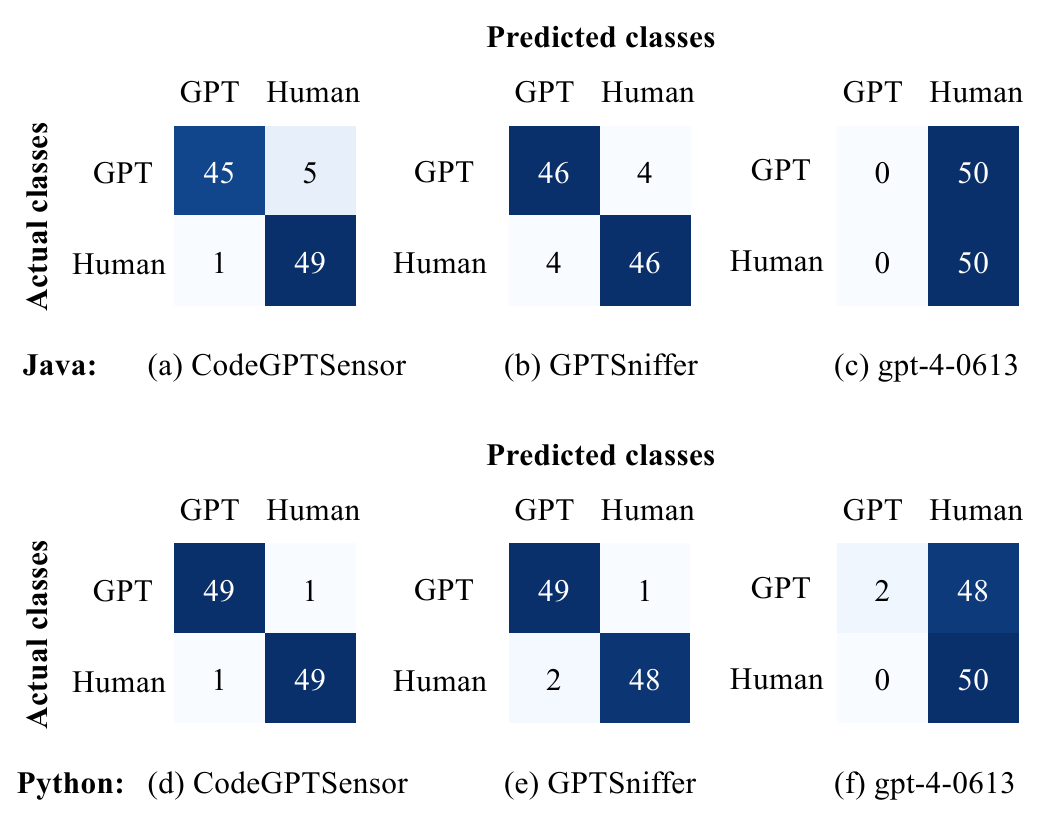}
    }
    \caption{Confusion Matrices of \toolname, GPTSniffer, and gpt-4-0613 for detecting ChatGPT-generated code on the sampled Java and Python subsets.}
    \label{fig:confusion}
    \Description{}
\end{figure*}

\addvspace{0.5em}
The responses of GPT-4 were collected through OpenAI API using the ``gpt-4-0613'' model in August 2024. For a fair comparison, we also evaluated \toolname and the best-performing baseline model GPTSniffer on the same subset using the metrics introduced in Section~\ref{sec:eval_metrics}.
We converted the ``logprob'' of the token ``True'' returned by GPT-4's API into the probability for the positive class, and the ``logprob'' of the token ``False'' into the probability for the negative class. The normalized probability score for the positive class was then used to calculate the AUC scores. 
The experimental results are shown in Table~\ref{tab:rq3_gpt4}. 
The results indicate that GPT-4 exhibits significantly lower performance in detecting ChatGPT-generated code compared to GPTSniffer and \toolname across almost all metrics for both Python and Java. 
As shown in Figure~\ref{fig:confusion}, GPT-4 exhibits a tendency to label code snippets as human-written, resulting in a notably poor recall. 
In contrast, GPTSniffer and \toolname achieve high scores across all evaluation metrics on the same subsets, demonstrating their strong abilities to distinguish between human-written and ChatGPT-generated code.

\intuition{
\textbf{Answer to RQ-3}: 
According to our evaluation, existing commercial AIGC detectors, zero-shot detectors, and even GPT-4 perform poorly in detecting ChatGPT-generated code.
Fine-tuning can substantially improve the performance of pre-trained model-based detectors. \toolname is more effective in distinguishing ChatGPT-generated code and less affected by code comments than all the selected baselines. 
}

%% file: sections/discussion.tex
\section{Discussion}
\label{sec:discussion}

\subsection{Comparison of Different Encoding Models}
\label{sec:code_encoder}

In this subsection, we compare the performance of \toolname using different encoding models in detecting LLM-generated code. 
Encoder-only models for source code pre-train a bidirectional Transformer encoder, which can greatly enhance the performance of code-related understanding tasks. CodeBERT~\cite{feng2020codebert}, which is pre-trained on NL-PL pairs using MLM and replaced token detection, and GraphCodeBERT~\cite{guo2020graphcodebert}, which leverages data flow to enhance code representation, are of the state-of-the-art encoder-only models for code.
UniXcoder~\cite{guo2022unixcoder}, a unified pre-trained model that supports both understanding and generation tasks, leverages code comment and AST to enhance code representation and introduces two new pre-training tasks to learn embeddings that can represent the semantics of code fragments. UniXcoder has been proven to outperform CodeBERT and GraphCodeBERT in code understanding tasks like clone detection and code search~\cite{guo2022unixcoder}.
Focusing on the task of detecting LLM-generated code, we compare the performance of \toolname using CodeBERT, GraphCodeBERT, and UniXcoder as the base encoding models, respectively.

\begin{table*}[t]
  \centering
  \caption{Detection results of \toolname with different encoding models. For simplicity, only comprehensive metrics such as F1-score and AUC are listed.}
  \resizebox{0.95\linewidth}{!}{
    % Table generated by Excel2LaTeX from sheet '【major】'
    \begin{tabular}{c|l|cc|cc|cc|cc}
    \toprule
    \multirow{2}[4]{*}{\textbf{Method}} & \multicolumn{1}{c|}{\multirow{2}[4]{*}{\textbf{Encoder}}} & \multicolumn{2}{c|}{\textbf{Python}} & \multicolumn{2}{c|}{\textbf{Java}} & \multicolumn{2}{c|}{\textbf{Python-no-comment}} & \multicolumn{2}{c}{\textbf{Java-no-comment}} \\
    \cmidrule{3-10}      &       & \textbf{F1-score} & \textbf{AUC} & \textbf{F1-score} & \textbf{AUC} & \multicolumn{1}{c}{\textbf{F1-score}} & \multicolumn{1}{c|}{\textbf{AUC}} & \multicolumn{1}{c}{\textbf{F1-score}} & \multicolumn{1}{c}{\textbf{AUC}} \\
    \midrule
    \multirow{3}[2]{*}{\toolname} & CodeBERT & 0.988  & \textbf{0.999 } & 0.952  & 0.989  & 0.961  & 0.993  & 0.945  & 0.987  \\
          & GraphCodeBERT & 0.966  & 0.994  & 0.944  & 0.989  & 0.964  & 0.994  & 0.944  & 0.989  \\
          & UniXcoder & \textbf{0.992 } & \textbf{0.999 } & \textbf{0.968 } & \textbf{0.995 } & \textbf{0.973 } & \textbf{0.995 } & \textbf{0.964 } & \textbf{0.994 } \\
    \bottomrule
    \end{tabular}%
  }
  \label{tab:encoder}%
\end{table*}%

 Experimental results are presented in Table~\ref{tab:encoder}.  
Overall, \toolname using UniXcoder achieves the best results in the task of LLM-generated code detection. 
\toolname using GraphCodeBERT incurs the highest time and space costs because it requires parsing Data Flow Graphs (DFG). However, this additional information hardly contributes to improving performance in detecting code generated by LLMs.
% We leave it for future works to evaluate the performance of smaller code pre-trained models in detecting LLM-generated code.

% \addvspace{0.2em}
\intuition{
\textbf{Finding of Section~\ref{sec:code_encoder}}: 
Among the selected encoding models, UniXcoder is the best one for building \toolname to detect LLM-generated code.
}

\subsection{Effectiveness of Contrastive Learning}
\label{sec:ablation}
In this subsection, we evaluate the effectiveness of contrastive learning. We compare the performance of \toolname and \toolname without the contrastive learning component (denoted as w/o Contrast).

\begin{table*}[t]
  \centering
  \caption{Detection results of \toolname compared with its variant. For simplicity, only comprehensive metrics such as F1-score and AUC are listed.}
  \resizebox{0.85\linewidth}{!}{
  {
    % Table generated by Excel2LaTeX from sheet '【major】'
    \begin{tabular}{l|cc|cc|cc|cc}
    \toprule
    \multicolumn{1}{c|}{\multirow{2}[4]{*}{\textbf{Method}}} & \multicolumn{2}{c|}{\textbf{Python}} & \multicolumn{2}{c|}{\textbf{Java}} & \multicolumn{2}{c|}{\textbf{Python-no-comment}} & \multicolumn{2}{c}{\textbf{Java-no-comment}} \\
    \cmidrule{2-9}      & \textbf{F1-score} & \textbf{AUC} & \textbf{F1-score} & \textbf{AUC} & \multicolumn{1}{c}{\textbf{F1-score}} & \multicolumn{1}{c|}{\textbf{AUC}} & \multicolumn{1}{c}{\textbf{F1-score}} & \multicolumn{1}{c}{\textbf{AUC}} \\ 
    \midrule
    \toolname & \textbf{0.992 } & \textbf{0.999 } & \textbf{0.968 } & \textbf{0.995 } & \textbf{0.973 } & \textbf{0.995 } & \textbf{0.964 } & \textbf{0.994 } \\
    - \textit{w/o Contrast} & 0.986  & \textbf{0.999 } & 0.961  & 0.993  & 0.964  & \textbf{0.995 } & 0.954  & 0.991  \\
    \bottomrule
    \end{tabular}%
   }
  }
  \label{tab:ablation}%
\end{table*}%

Experimental results are shown in Table~\ref{tab:ablation}. According to the results, we can observe that \toolname outperforms its \textit{w/o Contrast} version, especially in terms of F1-score (0.986$\rightarrow$0.992 in the Python test set, 0.961$\rightarrow$0.968 in the Java test set, 0.964$\rightarrow$0.973 in the Python-no-comment test set, and 0.954$\rightarrow$0.964 in the Java-no-comment test set). This suggests that utilizing contrastive learning can enhance the effectiveness of UniXcoder in distinguishing between human-written and ChatGPT-generated code.

\intuition{
\textbf{Finding of Section~\ref{sec:ablation}}: 
Employing contrastive learning can improve the pre-trained model's ability to distinguish between human-written and ChatGPT-generated code.
}

\subsection{Effectiveness of \toolname on Other Data Sources}
\label{sec:dataleak}

In this subsection, we additionally evaluate the performance of \toolname compared with other training-based methods on a dataset beyond the scope of CodeSearchNet to eliminate the possible impact of data leakage and to evaluate the generalizability of \toolname.
Specifically, we employ the $D_\alpha\text{-}C_8$ dataset\footnote{\url{https://github.com/MDEGroup/GPTSniffer/tree/master/DATASETS/RQ1/C8}} as used in the work of GPTSniffer~\cite{NGUYEN2024112059} and adhere to the same train/test data split. The $D_\alpha\text{-}C_8$ dataset consists of 1,484 human-written and ChatGPT-generated Java code snippets, where human-written ones mainly come from problem implementations from a book on Java programming~\cite{javasolution}, which are not included in the CodeSearchNet dataset that we use to build \datasetname. 

\begin{table*}[t]
  \centering
  \caption{Detection results of different training-based methods on the $D_\alpha\text{-}C_8$ test set.}
  \resizebox{0.75\linewidth}{!}{
  {
    \begin{tabular}{cl|ccccc}
    \toprule
    \multicolumn{2}{c|}{\multirow{2}[3]{*}{\textbf{Method}}} & \multicolumn{5}{c}{$\mathbf{D_\alpha\text{-}C_8}$} \\
    \cmidrule{3-7}\multicolumn{2}{c|}{} & \multicolumn{1}{c}{\textbf{Accuracy}} & \multicolumn{1}{c}{\textbf{Recall}} & \multicolumn{1}{c}{\textbf{Precision}} & \multicolumn{1}{c}{\textbf{F1-score}} & \multicolumn{1}{c}{\textbf{AUC}} \\
    \midrule
    {\multirow{3}{2.5cm}{\centering {RoBERTa-based}}} & RoBERTa-single* & 0.831  & 0.753  & \textbf{0.986 } & 0.854  & 0.958  \\
    \multicolumn{1}{c}{} & RoBERTa-QA* & 0.898  & 0.883  & 0.919  & 0.901  & 0.957  \\
    \multicolumn{1}{c}{} & GPT2-Detector* & 0.902  & 0.852  & 0.973  & 0.909  & 0.972  \\
    \midrule
    CodeBERT-based & GPTSniffer & 0.929  & \textbf{0.932 } & 0.926  & 0.929  & 0.972  \\
    \midrule \rowcolor{lightgray}
    Ours  & \toolname & \textbf{0.936 } & 0.912  & 0.957  & \textbf{0.934 } & \textbf{0.976 } \\
    \bottomrule
    \end{tabular}%
   }
  }
  \label{tab:data_leak}%
\end{table*}%

Table \ref{tab:data_leak} shows the detection results of \toolname compared with the fine-tuned RoBERTa-based and CodeBERT-based methods. According to the results, \toolname achieves the optimal performance in terms of Accuracy (0.936), F1-score (0.934), and AUC (0.976). 
This demonstrates that \toolname can effectively differentiate between human-written and ChatGPT-generated code in datasets beyond the CodeSearchNet corpus.

\addvspace{0.2em}
\intuition{
\textbf{Finding of Section~\ref{sec:dataleak}}: 
On a dataset that differs in data source and scale from \datasetname (i.e., $D_\alpha\text{-}C_8$), \toolname also achieves the best performance in detecting LLM-generated code, indicating that \toolname has good generalizability.
}

% -------------------------------------
\subsection{Influence of Randomness}
\label{sec:cross_validation}

In this paper, we randomly split all human-written and ChatGPT-generated code pairs into training, validation, and test sets at a ratio of 8:1:1, following previous works~\cite{hin2022linevd, li2021vulnerability}. Since our dataset is balanced and large in scale (i.e., consisting of 288K Python and 222K Java code pairs), the randomness in data partitioning will not have a significant impact on model evaluation. 
The same training/validation/test sets are used for all evaluated methods, which are all balanced datasets. Besides, we set the same random seed during all experiments.

In machine learning, particularly when data is insufficient, 10-fold cross-validation is a widely used method to assess a model's generalization ability~\cite{weiss1991computer}. 
However, applying 10-fold cross-validation to a large dataset is time-consuming.
To evaluate the performance of \toolname compared against the selected training-based methods using 10-fold cross-validation, we randomly select 20\% of the code pairs from the full dataset to create a representative subset, referred to as \datasetname-small. 
To perform 10-fold cross-validation, \datasetname-small is divided into 10 equally sized folds. In each iteration, one fold is held out for validation while the remaining nine folds are used for training. 

\begin{table*}[t]
  \centering
  \caption{Average performance metrics across 10-fold cross-validation on \datasetname-small.}
  \resizebox{\linewidth}{!}{
  {
    % Table generated by Excel2LaTeX from sheet '【major】'
    \begin{tabular}{cl|ccccc|ccccc}
    \toprule
    \multicolumn{2}{c|}{\multirow{2}[4]{*}{\textbf{Method}}} & \multicolumn{5}{c|}{\textbf{Python}}  & \multicolumn{5}{c}{\textbf{Java}} \\
    \cmidrule{3-12}\multicolumn{2}{c|}{} & \textbf{Accuracy} & \textbf{Recall} & \textbf{Precision} & \textbf{F1-score} & \textbf{AUC} & \textbf{Accuracy} & \textbf{Recall} & \textbf{Precision} & \textbf{F1-score} & \textbf{AUC} \\
    \midrule
    \multirow{3}[2]{*}{RoBERTa-based} & RoBERTa-single & 0.977  & 0.969  & \textbf{0.984 } & 0.977  & 0.997  & 0.875  & 0.823  & 0.826  & 0.824  & 0.938  \\
    \multicolumn{1}{c}{} & RoBERTa-QA & 0.981  & 0.984  & 0.979  & 0.981  & 0.998  & 0.935  & 0.951  & 0.922  & 0.936  & 0.984  \\
    \multicolumn{1}{c}{} & GPT2-Detector & 0.982  & 0.985  & 0.979  & 0.982  & 0.998  & 0.938  & 0.953  & 0.926  & 0.939  & 0.986  \\
    \midrule
    CodeBERT-based & GPTSniffer & 0.984  & 0.990  & 0.979  & 0.984  & \textbf{0.999 } & 0.947  & 0.960  & 0.935  & 0.947  & 0.989  \\
    \midrule \rowcolor{lightgray}
    Ours  & \toolname & \textbf{0.987 } & \textbf{0.991 } & 0.983  & \textbf{0.987 } & \textbf{0.999 } & \textbf{0.957 } & \textbf{0.961 } & \textbf{0.953 } & \textbf{0.957 } & \textbf{0.992 } \\
    \bottomrule
    \end{tabular}%
   }
  }
  \label{tab:10-fold}%
\end{table*}%

The average performance metrics for each model over the 10 iterations of training and validation are presented in Table~\ref{tab:10-fold}. 
From the results, we can observe that: 1) the RoBERTa-based models show variability in performance, with notable improvements from RoBERTa-single, RoBERTa-QA, to GPT2-Detector, but generally underperform compared to GPTSniffer and \toolname, especially in the Java datasets; 
2) GPTsniffer performs well, particularly in the Python datasets, but is still outperformed by \toolname in most metrics; 
3) \toolname outperforms other models on both Python and Java datasets in almost all metrics. 
These results indicate that \toolname is the most effective and reliable method for detecting LLM-generated code, demonstrating high generalization ability and robust performance.

\addvspace{0.2em}
\intuition{
\textbf{Finding of Section~\ref{sec:cross_validation}}: 
The results of the 10-fold cross-validation indicate that
among the evaluated training-based methods, \toolname is the most effective and reliable approach for detecting LLM-generated code.
}

% -------------------------------------
\subsection{Influence of Code Quality}
\label{sec:code_quality_v2}
% *************************
In this subsection, we explore the potential influence of code quality on \toolname's performance. 
To this end, we conducted an additional experiment to investigate whether code quality, particularly in terms of whether the code can pass compilation, introduces any significant bias in \toolname's identification process.
To minimize the risk of data leakage, we collected code snippets from three non-forked repositories created after 2022 on GitHub. 
The selected projects (i.e., HertzBeat~\cite{hertzbeat}, Fury~\cite{fury}, and OpenKoda~\cite{openkoda}) are all Maven projects that support compilation testing.
We used JavaParser~\cite{JavaParser} to extract code snippets, each consisting of a function accompanied by a docstring.
We then used ChatGPT (``gpt-3.5-turbo'') to generate the corresponding code based on the docstring and function signature using the following prompt:

\addvspace{0.3em}
\promptbox{Here is a function signature and its description. \\
Function description: \{docstring\} \\
Function signature: \{signature\} \\
Please generate the complete function implementation in Java. Do not write explanations, just output the code.}

In total, we collected 376 pairs of human-written and ChatGPT-generated functions for the compilation test. 
Note that since developers typically conduct necessary testing before uploading code to GitHub repositories, all 376 human-written functions we collected successfully passed compilation.
For each ChatGPT-generated function, we first reset and cleaned the project repository directory using Git's \textit{``reset''} and \textit{``clean''} commands. 
We then inserted the generated function into the original class and ran Maven's \textit{``compile''} command, capturing the output and any error messages from the compilation process.
As a result, 135 ChatGPT-generated functions passed compilation, yielding a compilation success rate of 35.9\%.
While this rate is low, it is important to clarify that the primary focus of this paper is on identifying the origin of the code  (i.e., whether it is human-written or ChatGPT-generated) rather than on ensuring or assessing the quality of ChatGPT-generated code. 
Future work on constructing datasets for detecting LLM-generated code in industrial settings may benefit from incorporating additional contextual information during code generation to improve code quality.

% *************************

\begin{table*}[t]
  \centering
  \caption{{Results of whether the collected functions can pass compilation, as well as the detection performance of \toolname. ``\#Sample'' stands for the number of functions.}}
  \resizebox{0.6\linewidth}{!}{
    {
        % Table generated by Excel2LaTeX from sheet 'metrics'
        \begin{tabular}{lccc}
        \toprule
        Source & Compilation Status & \#Samples & Accuracy \\
        \midrule
        Human-written & Passed & 135   & 0.807  \\
        ChatGPT-generated   & Passed & 135   & 0.844  \\
        \midrule
        Human-written & Passed & 241   & 0.817  \\
        ChatGPT-generated   & Failed & 241   & 0.830  \\
        \bottomrule
        \end{tabular}%
    }
  }
  \label{tab:code_quality_v2}%
\end{table*}%

We then evaluated \toolname's accuracy on both compilable and non-compilable code within each category. 
The experimental results are summarized in Table~\ref{tab:code_quality_v2}. 
Additionally, we performed a two-proportion Z-test using Python's \textit{proportions\_ztest}~\cite{ztest} to evaluate whether there is a statistically significant difference in \toolname's detection accuracy across different sets of code.
According to Table~\ref{tab:code_quality_v2}, for human-written code, \toolname achieved an accuracy of 0.807 on 135 compilable samples and 0.817 on another 241 compilable samples. The Z-test yielded a Z-statistic of -0.314 and a $p$-value of 0.754 (> 0.05), indicating no significant difference.
For ChatGPT-generated code, \toolname's accuracy was 0.844 on 135 compilable samples and 0.830 on 241 non-compilable samples. The Z-test yielded a Z-statistic of 0.178 and a $p$-value of 0.858 (> 0.05), also indicating no significant difference.
The results demonstrate that \toolname's performance is not significantly impacted by whether the code is compilable.
This suggests that \toolname does not rely solely on compilability as a heuristic for identification but instead effectively distinguishes between human-written and ChatGPT-generated code based on more subtle and intrinsic characteristics.

\addvspace{0.5em}
\intuition{
\textbf{Finding of Section~\ref{sec:code_quality_v2}}: 
Based on the experimental results and statistical analysis, \toolname's performance is not significantly influenced by whether the code is compilable. 
}

% -------------------------------------
\subsection{Comparison Between \toolname and GPTSniffer}
\label{sec:compare_advantage}
In this subsection, we explore the differences between \toolname and GPTSniffer from a qualitative perspective. 
While both \toolname and GPTSniffer achieve high metric values in detecting ChatGPT-generated code, as detailed in Table~\ref{tab:with_comment} and Table~\ref{tab:result_no_comment}, they differ in their design philosophies and unique features. 
\toolname, for instance, employs contrastive learning to fine-tune UniXcoder for enhanced class-separation code representations, while GPTSniffer simply fine-tunes CodeBERT for binary classification.

\begin{figure}[t]
\centerline{
    \includegraphics[width=.92\linewidth]{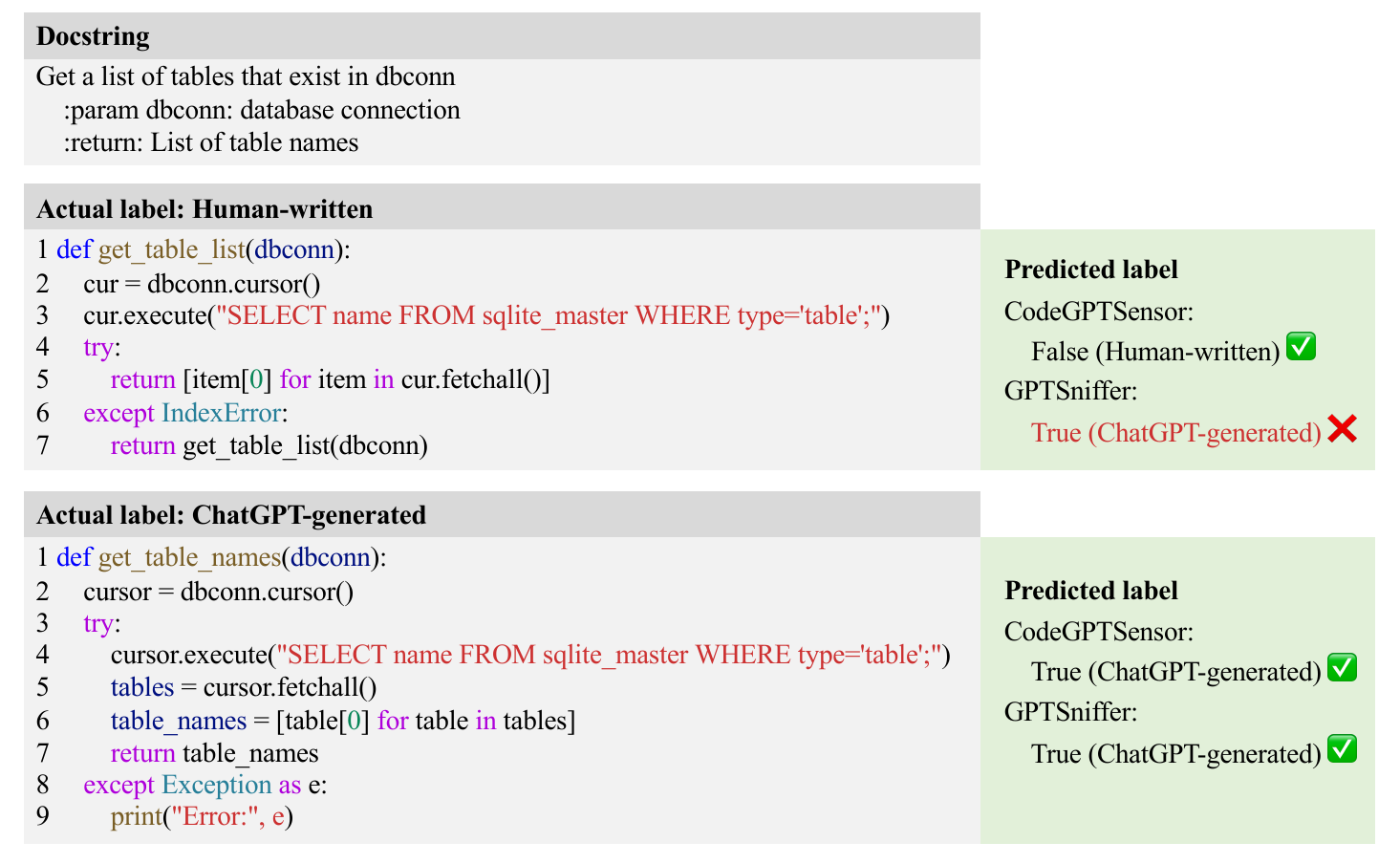}
    }
    \caption{{An example pair of highly similar human-written and ChatGPT-generated Python functions from \datasetname test set with index gp125095~\cite{diffexample}, where GPTSniffer misclassified the human-written function as ChatGPT-generated while \toolname correctly classified both functions.}} 
    \Description{}
    \label{fig:diff_example}
\end{figure}

Through qualitative analysis, we found that \toolname is more accurate in distinguishing between GPT-generated code and human-written code in cases when the two are highly similar. 
Figure~\ref{fig:diff_example} shows an example pair of functions with similar wording but different code structures.
While both \toolname and GPTSniffer correctly classified the ChatGPT-generated function, GPTSniffer misclassified the human-written function as ChatGPT-generated.
Comparing the two functions, the ChatGPT-generated one defines ``tables'' and ``table\_names'' to store intermediate results and uses a common error-handling method of printing error information.
In contrast, the human-written code uses ``cur'' as an abbreviation for ``cursor'' and employs a more complex and non-standard approach of error-handling that recursively calls ``get\_table\_list''. 
The atypical recursive error handling reflects a level of creativity and problem-solving more likely to be seen in human-written code, which \toolname successfully captured. 
\toolname's ability to correctly identify the human-written function indicates that it is more sensitive to the nuances of human programming styles, making it more robust and reliable in differentiating between human-written code and ChatGPT-generated code compared to GPTSniffer.

%% file: sections/threats_to_validate.tex
\section{Threats to Validate}

\noindent
\textbf{Threats to Internal Validity} mainly correspond to the potential mistakes in the implementation of our approach and other baselines. 
To minimize such threats, we not only implement these approaches by pair programming but also directly use the original source code from the GitHub or Hugging Face repositories shared by corresponding authors.
Besides, we use the same hyper-parameters as the original papers. 
We also carefully review the experimental scripts to ensure their correctness.

\noindent
\textbf{Threats to External Validity} may correspond to the generalizability of this study.
To mitigate this threat, we built the dataset by reusing the data from CodeSearchNet~\cite{husain2019codesearchnet}, which is collected from large amounts of different real-world projects hosted on GitHub.
This study is still limited in the following aspects: (1) the programming languages used in the studied functions are limited to Java and Python, and functions written by other popular programming languages (e.g., JavaScript or C/C++) have not been considered; 
(2) all the studied functions are collected from open-source projects, and the performance of \toolname on commercial projects is unknown; 
(3) ChatGPT is closed-source, which makes it impossible to dig into the internal workings of ChatGPT in code generation. 
Besides, ChatGPT is also continuously evolving. Our experiments only reflect the characteristics of code generated by a specific version of ChatGPT (i.e., gpt-3.5-turbo, April 2023) based on our designed prompt.
We plan to collect code snippets of other programming languages from more recent projects and evaluate more types of LLMs for code generation, especially open-sourced LLMs, in future work.

\noindent
\textbf{Threats to Construct Validity} mainly correspond to the performance metrics in our evaluations.
To minimize such threats, we consider a few widely used performance metrics to compare the detection performance between our proposed approach \toolname and baselines.
In particular, we totally consider five performance metrics 
(i.e., Accuracy, Precision, Recall, F1-score, and AUC).

%% file: sections/related_work.tex
\section{Related Work}
\label{sec:related_work}

\subsection{AI in Software Engineering}

There are numerous tasks in software engineering, such as defect prediction~\cite{ni2022best}, defect localization~\cite{li2022fault}, vulnerability detection~\cite{zhou2019devign}, and code search~\cite{liu2023graphsearchnet}, where novel approaches based on AI, particularly recent neural network-based approaches, have made significant progress.
Various types of neural network architectures have been applied, including CNN~\cite{lecun2015deep}, GRU~\cite{cho2014properties}, LSTM~\cite{hochreiter1997long}, Graph Neural Networks~\cite{li2015gated}, and Transformer~\cite{vaswani2017attention}. 
In recent years, transformer-based models, in particular, have been widely studied and applied, which can be divided into three architectural types: encoder-only, encoder-decoder, and decoder-only.

As for the encoder-only models, CodeBERT~\cite{feng2020codebert}, GraphCodeBERT~\cite{guo2020graphcodebert}, and UniXcoder~\cite{guo2022unixcoder} are the representative ones, which are pre-trained as general models on code-related data and can be fine-tuned for downstream tasks to achieve superior performance.
As for the encoder-decoder models, there are also some works (e.g., PLBART~\cite{ahmad2021unified} and CodeT5~\cite{wang2021codet5}) proposed to enhance the model capacity.
These pre-trained models have achieved significant improvements in many important software engineering tasks.
However, they do not bring researchers or participants with exciting results.
Then, another architecture (decoder-only) attracts a small portion of people's attention.
GPT~\cite{radford2018improving} is the most representative model, potentially bringing the large language model into practical usage.
First, the CodeX model (i.e., GPT variant)~\cite{chen2021evaluating}, trained on publicly available code from GitHub, is released by OpenAI.
Following that, another powerful tool named Copilot is trained, which is an advanced version of CodeX and can bring developers much help with coding.
Recently, the ChatGPT model has attracted the widest attention from the world, which is the successor of the large language model InstructGPT~\cite{ouyang2022training} with a dialog interface that is fine-tuned using the Reinforcement Learning with Human Feedback (RLHF) approach~\cite{christiano2017deep}.
Benefiting from its massive learned knowledge and powerful conversation capabilities, ChatGPT can generate accurate responses across various domains.
Therefore, ChatGPT is used to assist many software engineering tasks, such as code generation~\cite{nair2023generating}, vulnerability detection~\cite{cheshkov2023evaluation}, and defect repair~\cite{surameery2023use,xia2023keep}.
For example, Xia et al.~\cite{xia2023keep} proposed a novel method, ChatRepair, to evaluate the ability of ChatGPT on bug fixing with the dataset of Defects4j, and their results show that ChatRepair fixes 162 out of 337 bus for \$0.42 each, which outperforms the state-of-the-art approaches.
In this work, we aim to explore how to distinguish the code generated by ChatGPT.

\subsection{LLM-generated Content Detection}
Large language models are continuously iterated, which leads to dramatically improved performance on many language-related tasks and the ability to generate convincing text~\cite{choi2023chatgpt,zhang2022opt,mitchell2023detectgpt}.
The surprisingly strong capabilities of ChatGPT have raised many interests and concerns.
For example, people are either curious about how close ChatGPT is to human experts or are worried about the potential risks brought by LLMs like ChatGPT.
Therefore, to ensure the responsible and ethical use of LLM-generated content, it is necessary to propose approaches to identify whether a given piece of content is generated by LLMs.
Yang et al.~\cite{yang2023survey} summarize existing methods for LLM-generated content detection into three categories: 1) Training-based methods ~\cite{chatgptzero, aitextclassifier, zhan2023g3detector, chen2023gpt, yu2023gpt, liu2022coco, wu2023llmdet, tian2023multiscale, hu2024radar}, which usually involve fine-tuning a pre-trained language model on a constructed dataset of both human-written and LLM-generated content;
2) Zero-shot methods~\cite{mireshghallah2023smaller, krishna2024paraphrasing, mitchell2023detectgpt, yang2023dna}, which utilize intrinsic properties of typical LLMs, such as probability curve~\cite{mitchell2023detectgpt} and N-gram divergence~\cite{yang2023dna}, for self-detection; 
3) Watermarking involves concealing information within the generated text, which enables the identification of the text's source.
Our proposed approach falls within the category of training-based methods. Diverging from existing models that primarily focus on detecting text generated by LLMs, utilizing data sources such as Wikipedia~\cite{guo2023hc3} and student essays~\cite{verma2023ghostbuster}, our work concentrates on identifying program code generated by LLMs.

% --- empirical study
In the domain of detecting code generated by LLMs, Pan et al.~\cite{pan2024assessing} conduct an empirical study using a dataset derived from fundamental Python programming problems to evaluate the effectiveness of several AIGC detectors~\cite{chatgptzero, Sapling, gpt2detector, mitchell2023detectgpt, gehrmann2019gltr} in detecting AI-generated code. Their findings reveal the inadequate performance of these detectors in identifying AI-generated code, highlighting the urgent need for more reliable detection methods, particularly in educational contexts.
% --- GPTSniffer
The recently released work by Nguyen et al.~\cite{NGUYEN2024112059}, \textit{independent and concurrent to our work}, proposes a CodeBERT-based classifier named GPTSniffer that is similar to our idea. Our work is different in the following aspects: 1) Nguyen et al. focus on assessing the impact of different preprocessing settings on prediction performance instead of developing the best AI-generated code detector, while our work proposes a novel detection approach that fine-tunes UniXcoder with contrastive learning and achieves optimal prediction performance; 
2) The dataset used in Nguyen et al.'s work is limited in scale, consisting of only 1.5K code snippets derived from common programming tasks or exercises of a particular book~\cite{liang2003introduction}. In contrast, our dataset consists of 550K pairs of human-written and ChatGPT-generated code sourced from GitHub, which more closely replicates real-world software development scenarios;
3) Apart from evaluation experiments, we also perform a qualitative analysis to discover certain characteristics of ChatGPT-generated code.
% --- zero-shot
Different from our training-based method, Yang et al.~\cite{yang2023zero} concentrate on the zero-shot detection of LLM-generated code and introduce DetectGPT4Code, a method utilizing a small proxy model to approximate the logits on the conditional probability curve. 
% --- watermarking
Inspired by previous text watermarking techniques~\cite{kirchenbauer2023watermark}, Lee et al.~\cite{lee2023wrote} propose watermarking for code generation, which involves selectively injecting watermarks to tokens with entropy higher than a predefined threshold, to enhance the detectability and generated code quality.

%% file: sections/conclusion.tex
\section{Conclusion}

In this paper, to facilitate the analysis of the distinctive characteristics of human-written and ChatGPT-generated code, we curate a large-scale dataset, \datasetname, consisting of 288K Python and 222K Java pairs of human-written and ChatGPT-generated code snippets. 
Our manual evaluations and code complexity analysis on the \datasetname dataset shed light on the differences between human-written and ChatGPT-generated code.
We propose the \toolname model to effectively detect ChatGPT-generated code, which leverages the contrastive learning framework with a semantic encoder built with UniXcoder.
Our experiments demonstrate that our proposed model, \toolname, is more effective in detecting ChatGPT-generated code than existing AIGC detection tools. 
This advancement can potentially assist developers and educators in distinguishing between ChatGPT-generated and human-written code more easily and accurately.